\documentclass[twocolumn,twocolappendix,trackchanges,tighten]{aastex701}
\usepackage{graphicx}	
\usepackage{subfigure} 
\usepackage{amsmath}	
\usepackage{orcidlink}
\usepackage{longtable}
\usepackage{multirow}
\usepackage{caption}
\usepackage{makecell}
\usepackage{upgreek}

\def\ol{\overline}

\def\bm{\boldsymbol}

\newcommand{\e}{\mathrm{e}}  
\newcommand{\di}{\mathrm{d}}  

\usepackage{commands}
\usepackage[normalem]{ulem} 
\usepackage{cancel} 

\usepackage{xparse} 

\NewDocumentCommand{\SGc}{O{} m}{\textcolor{cyan}{[SG comments#1: #2]}}

\definecolor{mypurple}{rgb}{0.7,0.3,0.8}
\renewcommand{\as}[1]{\textcolor{mypurple}{#1}} 

\begin{document}

\title[Radio Continuum from Evolving SFGs -- I]{Radio Continuum Emission from Evolving Star-Forming Galaxies~--~I. Correlations Involving the Total Synchrotron Luminosity}

\author[orcid=0009-0009-4404-1684,sname='Ghosh']{Sukanta Ghosh}
\affiliation{National Institute of Science Education and Research, Bhubaneswar 752050, India}
\affiliation{Homi Bhabha National Institute, Training School Complex, Anushakti Nagar, Mumbai 40094, India}
\email[show]{sghosh.astrophy@gmail.com}  

\author[orcid=0000-0003-4935-5550,sname='Chamandy']{Luke Chamandy}
\affiliation{National Institute of Science Education and Research, Bhubaneswar 752050, India}
\affiliation{Homi Bhabha National Institute, Training School Complex, Anushakti Nagar, Mumbai 40094, India}
\email[show]{lchamandy@niser.ac.in}  
\correspondingauthor{Luke Chamandy}

\author[orcid=0000-0003-1494-7177,sname='Jose']{Charles Jose}
\affiliation{Department of Physics, CUSAT, Cochin, 682022, India}
\email{charles.jose@cusat.ac.in}  

\author[orcid=0000-0001-6200-4304,sname='Shukurov']{Anvar Shukurov}
\affiliation{School of Mathematics, Statistics and Physics, Newcastle University, Newcastle upon Tyne, NE1 7RU, UK}
\email{anvar.shukurov@ncl.ac.uk}  

\author[orcid=0000-0002-3860-0525,sname='Rodrigues']{Luiz Felippe S. Rodrigues}
\affiliation{Antonie, Uitmeentsestraat 19, 6987 CX Giesbeek, Netherlands}
\email{luiz.rodrigues@antonie.ai}  

\author[orcid=0000-0002-0377-0970,sname='Tabatabaei']{Fatemeh Tabatabaei}
\affiliation{School of Astronomy, Institute for Research in Fundamental Sciences (IPM), PO Box 19395-5531, Tehran, Iran}
\email{ftaba@ipm.ir}

\defcitealias{Rodrigues+19}{R19}
\defcitealias{Jose+24}{J24}
\defcitealias{Tabatabaei+16}{T16}
\defcitealias{Tabatabaei+17}{T17}
\defcitealias{Heesen+22}{H22}
\defcitealias{Hosseinirad+23}{H23}
\defcitealias{Lacey+16}{L16}

\begin{abstract}

Synchrotron radiation dominates the continuum emission of star-forming galaxies 
in the frequency range from a few MHz to about $30\GHz$. 
We model the total synchrotron emission of a large population of evolving star-forming galaxies 
using the semi-analytic galaxy formation model \textsc{galform} combined with the dynamo simulation code \textsc{magnetizer}.
Assuming local energy equipartition between cosmic rays and magnetic fields, we calculate the specific synchrotron luminosity $\Lspec$ for each simulated galaxy at various frequencies
and find strong positive correlations between $\Lspec$ and both the star formation rate ($\SFR$)
and characteristic galaxy rotation speed $V_{\rm rot}$ for redshifts up to $z\simeq 3$. 
At low redshifts, the turbulent magnetic field is found to dominate in the synchrotron luminosity, 
but the contribution of the large-scale magnetic field increases with redshift and becomes important for $z\gtrsim1$. 
The correlation between $\Lspec$ and $\SFR$ arises from the tight correlation between the disc gas mass $M\gas$ and $\SFR$,
and the correlation between $\Lspec$ and $V\rot$ is additionally a consequence 
of the stellar mass Tully--Fisher relation for main-sequence galaxies.
At low redshifts, the model predictions and observational data compiled for this work show remarkable agreement,
but a discrepancy arises at higher redshifts, 
where modelled SFR values are systematically smaller than those previously inferred from observations.
These theoretical models will aid the interpretation of next-generation radio surveys 
with the Square Kilometre Array and other telescopes. 
\end{abstract}



\section{Introduction}
Magnetic fields pervade the interstellar media of star-forming galaxies (SFGs) 
and have energy densities comparable to those of turbulence, 
making them dynamically important \citep{Beck+Wielebinski13,Klein+Fletcher15,Beck+19}.
Highly sensitive radio telescopes such as the Square Kilometre Array (SKA) and its precursors 
can probe the magnetic fields of statistical populations of SFGs by observing their synchrotron emission.
Theoretical physics-based models are needed to produce synthetic data 
that can be compared with current and future observational data, including for SFGs at high redshift.

The radio luminosity function of SFGs is one such data product that is now being modeled theoretically 
(\citealt{Jose+24}, hereafter \citetalias{Jose+24}; \citealt{Hansen+24}; \citealt{Prathap+25}).
Equally important is the study of the statistical relationships between observable properties, 
which can lead to the discovery of correlations, i.e., scaling relations. 
Early studies by \citet{Bell+03} demonstrate a tight correlation between $1.4\GHz$ radio luminosity 
and star formation rate (SFR) (computed using infrared luminosity) for nearby galaxies. 
Using deep GMRT observations, 
\cite{Garn+09} showed that a correlation between radio synchrotron luminosity and SFR is maintained up to $z\sim2$. 
This supports the use of radio continuum emission as a reliable, 
dust-unbiased tracer of star formation and cosmic star formation history across a wide range of redshifts \citep{An+26}.

\citet{Tabatabaei+16} (hereafter \citetalias{Tabatabaei+16}) 
discovered a scaling relation between the radio continuum specific luminosity at $4.8\GHz$ 
(expressed as the flux $S_{\rm{I}}$ at the fixed distance of $10\Mpc$) and $\SFR$, for isolated SFGs.
They also found correlations between $S_{\rm{I}}$ and the rotation speed 
in the flat part of the rotation curve, $V\rot$, 
between the linearly polarized specific flux at $4.8\GHz$, $S\PI$, and $\SFR$,
and between $S\PI$ and $V\rot$.
Additional specific flux density data were compiled by \citet{Tabatabaei+17} 
(hereafter \citetalias{Tabatabaei+17}), 
but they did not revisit these scaling relations, 
and then by \citet{Smith+21} and \citet{Heesen+22} (hereafter \citetalias{Heesen+22}),
who explored the correlation between specific radio luminosity $\Lspec$ and $\SFR$ at $150\MHz$ and $144\MHz$ using LOFAR data.
In this work, we combine the data sets of \citetalias{Tabatabaei+16}, \citetalias{Tabatabaei+17}, \citetalias{Heesen+22}, \citet{Heesen+26} and \citet{Bell+03} 
and perform a correlation analysis between $\Lspec$ and $\SFR$ and between $\Lspec$ and $V\rot$, 
at various frequencies.
The results are then compared with those of our theoretical model.

Synchrotron emission is produced by cosmic ray electrons moving helically about magnetic field lines, 
so theoretical models should account for both magnetic fields and cosmic rays in galaxies.
Galactic dynamo theory broadly explains certain basic observed properties 
of the interstellar magnetic fields of spiral galaxies, such as their global symmetries \citep{Beck+19},
but detailed dynamo models of specific galaxies have had mixed success \citep[e.g.,][]{Vaneck+15,Chamandy+16,Nazareth+25}.
The dynamo process can be conveniently divided into a fluctuation or small-scale dynamo 
that amplifies and sustains a turbulent smallƒscale contribution to the magnetic field,
and a mean-field or large-scale dynamo, which is responsible for the global large-scale field
\citep{Beck+96,Brandenburg+Subramanian05a,Shukurov05,Klein+Fletcher15,Beck+19,Shukurov+Subramanian21}.
The fluctuation dynamo also provides a weak large-scale field that seeds the mean-field dynamo, 
whereas turbulent tangling of the large-scale field enhances the small-scale field,
so the two types of dynamo are not completely separate.
Both types of dynamo have been found to operate simultaneously in high-resolution MHD simulations 
that simulate a localized region of a galaxy \citep{Gent+24},
but such models are not directly useful for studying large populations of galaxies.

Modeling the cosmic ray (CR) electron distribution (spatially, spectrally and temporally) 
is arguably as challenging as modelling the magnetic field \citep{Schlickeiser+02}.
To partially circumvent this difficulty it is commonly assumed that CRs and magnetic fields 
are in local energy equipartition,
which may be reasonable if the spatial resolution scale of observations 
is much larger than the turbulent correlation length 
(\citealt{Beck&Krause05}; \citealt{Seta+Beck19}; \S4.5 of \citealt{Shukurov+Subramanian21}).

One possible theoretical approach is to use 
lower resolution cosmological MHD simulations, 
which are now able to produce $\mkG$-strength magnetic fields in galaxies starting from weak seed fields. 
However, their resolution is still too low to realistically model the dynamo \citep[\S13.14.3 of][]{Shukurov+Subramanian21}. 
Simulations that rerun small parts of a cosmological MHD simulation with increased resolution (the so-called zoom simulations),
are better able to model certain dynamo processes~\citep{Pakmor+17,Pakmor+24}.
However, so far such methods can be used to simulate only a handful of galaxies at a time, which is insufficient for population studies.

Despite these challenges,
\citet{Hosseinirad+23} 
compared the magnetic fields from the cosmological MHD simulation \textsc{Illustris TNG50} to those they inferred 
from the \citetalias{Tabatabaei+16} observational data.
They explored correlations between the volume-averaged total, 
ordered, or turbulent magnetic field strength, on the one hand,
and $V\rot$, twice the stellar half-mass radius $r_{1/2}$, 
or characteristic angular rotation speed $\Omega$, on the other.
They restricted their analysis to centrals,
i.e.~galaxies located near the centers of their dark matter halos,
and found closer agreement for isolated galaxies in the simulation whose nearest neighbour is more than $2\Mpc$ away.
For this subset of simulated galaxies, 
the relation between the magnetic field components and $V\rot$ was found to agree rather well with observational inference,
with correlation coefficients and scaling law exponents generally consistent to within $\sim25$ per cent.
In particular, 
both the exponent and normalization of the scaling law found 
between the ordered component of the magnetic field and $V\rot$ 
are remarkably consistent in simulations and observations. 
However, the level of agreement for the relations involving $r_{1/2}$ and $\Omega$ is rather poor.
Another property for which \textsc{TNG50} simulations and observational inference from \citetalias{Tabatabaei+16} 
do not match is the normalization of the ratio of the ordered to the turbulent component of the magnetic field.
In the simulations, the strength of the ordered component is typically larger than that of the turbulent component, 
whereas in the observations the turbulent component tends to be stronger than the ordered component.
\citet{Hosseinirad+23} suggest possible explanations for this discrepancy.

Attempts have also been made to model the synchrotron luminosity of galaxy populations analytically 
or semi-analytically \citep[e.g.][]{Vollmer+22, Schober+23}. 
The main goal of those works was to investigate theoretically the 
well known correlation between far infrared (FIR) and radio emission observed for star-forming galaxies
\citep{Van_der_Kruit71,Yun+01,McCheyne+22}.
Such models 
have the advantage of simplicity
but models that incorporate the detailed physics of galaxy and magnetic field evolution
are also needed in order to make progress.

Our galaxy population code \textsc{magnetizer}
(\citealt{Rodrigues+15}; \citealt{Rodrigues+19}, hereafter \citetalias{Rodrigues+19}; 
\citealt{Rodrigues+Chamandy20}, \citealt{Jose+24}, hereafter \citetalias{Jose+24})
makes possible a hybrid approach in that it combines detailed semi-analytical modeling 
of galaxy formation and evolution with mean-field dynamo simulations,
and solves for galaxy properties as a function of both position within a galaxy and time, 
for a statistical population of sources.
In this work we simulate the total synchrotron emission from evolving galaxy populations 
and compare the predicted correlations between continuum radio emission and other star-forming galaxy properties with observations. 
We will explore the polarized component of the synchrotron emission and the FIR-radio correlation 
in future papers of this series.

The paper is organized as follows.
In Section~\ref{sec:model} we present the model.
Section~\ref{sec:data} summarizes the observational data used for comparison.
The main results of the model, 
as well as its comparison to the data, are presented in Section~\ref{sec:results},
and their implications discussed in Section~\ref{sec:discussion}. 
We summarize and conclude in Section~\ref{sec:conclusions}.

\section{Model}\label{sec:model}
The calculations presented in this work involve a three-stage process:
(i)~running a galaxy formation model to produce a sample of galaxies, 
(ii)~computing the magnetic fields in each galaxy of the sample, and 
(iii)~computing the total synchrotron specific luminosity (luminosity per unit frequency interval) for each galaxy. These stages are expounded on below;
further details can be found in \citetalias{Rodrigues+19} and \citetalias{Jose+24}.

\subsection{Galaxy formation model}
The first stage uses galaxy properties from the \textsc{galform} semi-analytic galaxy-formation model of \citet{Lacey+16} (hereafter \citetalias{Lacey+16}; 
see also \citealt{Cole+00,Baugh+05}), selecting data over the redshift range $0 \leq z \leq 6$. 
\textsc{galform} incorporates theoretical and empirical models for complex physical processes 
(star formation, supernova and AGN feedback, various dynamical processes, 
the evolution of gas, stars, and dust, etc.) 
on top of an $N$-body dark matter-only simulation \citep{Guo+2013}.
For this study, we used about $2\times10^5$ galaxies from the \citetalias{Lacey+16} version of \textsc{galform}, 
selected at $z=0$ (for selection criteria, see Section~\ref{sec:selection}),
except that we include a correction to the overall galaxy sizes (\citealt{Rodrigues+15}; \citetalias{Rodrigues+19})
in order to make the model more consistent with observational data. 
We obtained $45$ fixed-redshift snapshots of \textsc{galform} to use in the next stage of the model~--
the same snapshots used in \citetalias{Rodrigues+19}.
These are evenly spaced in the logarithm of the cosmological scale factor,
which corresponds to time intervals between $100$ and $300\Myr$.
Galaxy properties are 
kept constant between these snapshots 
(except for the large-scale magnetic field, as explained in Section~\ref{sec:LS}).
This is unlikely to be an important limitation since the results of our study 
do not change significantly when every second snapshot is excluded.

\subsection{Axisymmetric galaxy model with magnetic fields}\label{sec:magnetizer}
In the next stage, we run \textsc{magnetizer}, 
which produces an evolving spatial structure of each galaxy
in cylindrical polar coordinates $(r,\phi,Z)$ assuming axial symmetry ($\del/\del\phi=0$).
The galaxy rotation curve $V\rot(r)$ is calculated by modelling the galaxy as consisting of three main components: 
a thin stellar disc with an exponential surface mass density profile, 
a central bulge described by the \citet{Hernquist90} profile, and a dark matter halo 
following an adiabatically contracted Navarro--Frenk--White (NFW) density profile \citep{Navarro+1997}.  
The structure of the gaseous disc is obtained by assuming vertical hydrostatic equilibrium at each radius,
with the midplane pressure including contributions from turbulence, thermal gas, magnetic fields and cosmic rays.
This leads to a disc that is flared, i.e., its scale height $h\disc$ increases with $r$.
The maximum gas disc radius $(r\disc)$ is set to 
$2.7\,r_{1/2}$, with $r_{1/2}$ the half-mass radius for the baryonic mass.
We have verified that the results are not sensitive to the choice of $r\disc$, 
as long as it is sufficiently large compared to $r_{1/2}$.

In both \citetalias{Rodrigues+19} and \citetalias{Jose+24}, 
the total disc gas is divided into two components -- diffuse and molecular -- following the prescription of 
\citet{Blitz&Roso2004,Blitz&Roso2006},
and only the diffuse gas is used to calculate the magnetic fields.
However, the association of galactic magnetic fields with the diffuse gas alone is not 
obvious since each volume element in a multi-phase interstellar medium (ISM) can be transferred from one phase to another 
on a relatively short time scale.
Our fiducial model (\texttt{Fiducial}) makes use of \textit{all} of the gas in the disc 
for calculating the magnetic field strength rather than using just the diffuse component.
This improves the agreement of our results with observational data;
it is also simpler than and just as plausible as the diffuse gas assumption.

The root-mean-square turbulent speed $v\turb$ (comparable to the speed of energy-carrying eddies)
affects the hydrostatic balance and the strength of the magnetic field.
It is assumed to be uniform throughout a galaxy, but may vary between galaxies 
as a function of the global disc $\SFR$ (\citealt{Krumholz+18}; \citetalias{Jose+24}):
\begin{equation}\label{eq: sfr-vt}
    v\turb = \begin{cases}
v_0, & \text{if } \SFR \leq \SFR_0 \\
v_0\left( \dfrac{\SFR}{\SFR_0} \right)^{\rm{c}}, & \text{otherwise}.
\end{cases}
\end{equation}
where
$v_0 = 15\kms$, $\SFR_0 = 1\Msunyr$ and $c = 0.3$, 
as obtained by fitting the median of the observational data (see Appendix~\ref{app:v_turb-SFR} for details).

The magnetic field consists of two contributions:
(i)~an isotropic small-scale turbulent field $\vec{b}$ with the energy density set equal to a fixed fraction
of the local turbulent kinetic energy density and (ii)~a large-scale (mean) magnetic field $\meanv{B}(r)$ 
obtained by solving numerically a set of non-linear mean-field dynamo equations
assuming an $\alpha$$\Omega$-type dynamo.\footnote{\citetalias{Jose+24}
solved an $\alpha^2\Omega$-type dynamo rather than $\alpha\Omega$. 
However, we have checked that the $\alpha^2$ term has a small effect on the overall results.
For runs studied in this work (other than $\texttt{J24}$), 
we omit the $\alpha^2$ term because it is still unclear how  
to approximate this term in the no-$Z$ approximation for thin discs; see Section~\ref{sec:LS}.}
The non-linear effects of the mean-field dynamo 
are due to the magnetic helicity balance \citep{Brandenburg+Subramanian05a,Shukurov+Subramanian21}.
They lead to the saturation of the exponential growth of $\meanv{B}(r)$ 
at levels comparable to the equipartition field strength 
\begin{equation}\label{Beq}
  B\eq(r) = (4\uppi\rho)^{1/2}v\turb,
\end{equation}
where $\rho$ is the gas density.

\subsubsection{Small-scale magnetic field}\label{sec:SS}
The small-scale magnetic field is produced by the turbulent motion of the ISM 
through the small-scale or fluctuation dynamo and the tangling of the large-scale magnetic field. 
Its average statistical properties are sufficient to derive the total synchrotron emissivity.
The small-scale dynamo amplifies a weak seed magnetic field to a strength comparable to 
$B\eq$ within a timescale of the order of $10\Myr$ 
(comparable to the turbulent eddy turnover time of energy-carrying eddies; 
see, e.g., \citealt{Beck+94}, and \S6.7 and \S13.3 of \citealt{Shukurov+Subramanian21}). 
Since this timescale is much shorter than the timescale over which galaxies evolve, 
we assume that the small-scale field is saturated (i.e., remains in a statistically steady state) \textit{at all times},
with the rms strength given by
\begin{equation}\label{eq:1}
  b\rms = f_b\,B\eq. 
\end{equation}
The proportionality constant $f_b$ is set to 
$0.8$ in our fiducial model, 
which leads to a slightly better agreement with observations used in this paper than $f_b=1$ used in \citetalias{Jose+24}.
The existing estimates of $f_b$ vary between different models, remaining of order unity \citep[e.g.][]{Beck+19,Gent+24},
and $f_b = 0.8$ is consistent with the information available.

\subsubsection{Large-scale magnetic field component}\label{sec:LS}
Large-scale magnetic fields in galaxies are coherent on scales up to the system size \citep{Beck15c,Beck+19}.
Mean-field dynamo action amplifies the large-scale field with an $\e$-folding time comparable 
to the galactic rotation period $2\uppi/\Omega$ 
at approximately the radius where the galactic shear rate $|r\,\di\Omega/\di r|$ is the largest. 
The large-scale field evolution is simulated for each galaxy
using a finite difference code (6th-order in $r$ with the 3rd-order implicit Runge-Kutta time-stepping, 
\citealt{Brandenburg03}) after averaging over the $Z$-direction (across the disc thickness) 
using the `no-$Z$' approximation of 
dynamo theory
\citep{Subramanian+Mestel93,Moss95,Phillips01},
which has been shown to accurately recover key properties of $Z$-dependent solutions \citep[e.g.,][]{Chamandy+14b,Chamandy16}.
\textsc{magnetizer}
updates the dynamo input parameters for each \textsc{galform} snapshot
but the time resolution of the large-scale magnetic field evolution is much finer.

The saturated field strength of the large-scale magnetic field 
is proportional to the local equipartition field strength $B\eq(r)$. 
It also depends on the adjustable dimensionless parameter $R_\kappa$, 
which is equal to the ratio of the diffusivity of the magnetic contribution to the $\alpha$ effect ($\kappa$)
to that of the mean magnetic field (dominated by the turbulent diffusivity $\etat$).
In our model, the mean vertical outflow speed from the gaseous disc is assumed to be small 
as compared to the vertical turbulent diffusion speed 
($\mean{V}_Z\ll \etat/h\disc$, with $\etat$ the turbulent diffusivity of the mean magnetic field), 
which allows us to neglect outflows in the dynamo equations by setting $\mean{V}_Z=0$.
In this case, the mean magnetic field strength is approximately proportional to $R_\kappa^{1/2}$ 
in the saturated state \citep{Chamandy+14b}.

The large-scale seed magnetic field is obtained as the average of the small-scale turbulent magnetic field 
over a finite volume within the galactic disc. 
As explained in \citetalias{Rodrigues+19}, 
this gives a kiloparsec-scale random \textit{large-scale} magnetic field of order $10^{-3}\mkG$ in strength
(\S VII.14 of \citealt{Ruzmaikin+88}, \citealt{Poezd+93}, 
\citealt{Gent+24} and chapter~13 of \citealt{Shukurov+Subramanian21}).

When a galaxy is involved in a major merger 
(defined as a merger for which the ratio of the baryonic masses involved exceeds $0.3$),
or experiences a bar instability, the disc is 
destroyed and rapid star formation depletes the gas reservoir in \textsc{galform} \citepalias{Lacey+16}.  
\textsc{magnetizer} sets the large-scale field to zero 
when the disc gas mass and the disc radius fall below the threshold values of $10^4\Msun$ and $0.5\kpc$, respectively. 
After each \textsc{galform} snapshot, 
\textsc{magnetizer} rechecks these threshold values and if they are exceeded,
the mean-field dynamo is restarted with a seed field; 
otherwise the large-scale magnetic field remains zero.
Further details about the numerical implementation can be found in \citetalias{Rodrigues+19}. 

\subsubsection{3D galaxy model}
\textsc{magnetizer} computes the radial profiles of various galactic parameters 
(such as the gas density, scale height, and magnetic fields) at the midplane $Z=0$. 
The \textit{vertical} profiles of $\mbr(r,Z)$, $\mbp(r,Z)$, $b\rms(r,Z)$ and 
$\rho(r,Z)$ are assumed to be exponential, 
with the scale height equal to $2h\disc(r)$ for the magnetic fields and $h\disc$ for the gas density, 
as suggested by equation~\eqref{Beq}. 
\textsc{magnetizer} computes the $r$ and $\phi$ components of the large-scale magnetic field using the no-$Z$ approximation, 
which provides vertically averaged field strengths. 
The midplane values are $1/(1-e^{-1})$ times larger than those averages. 
The vertical magnetic field component $\mbz(r,Z)$ is reconstructed from $\Del\cdot\meanv{B}=0$
(see Appendix~\ref{app:Bz} for details).

\subsection{Synchrotron emission model}
The total intensity of the synchrotron emission is obtained by integrating the emissivity.
The synchrotron emissivity is proportional to the  product of the cosmic ray electron number density 
in the relevant energy range and approximately the square of the component 
of the magnetic field perpendicular to the line of sight (LoS).

The total synchrotron specific luminosity 
can be obtained by integrating the emissivity along the line of sight 
and over all lines of sight that pass through the galaxy.
When LoS effects like synchrotron self-absorption and free-free absorption are not important,
a volume integral can be used instead (which is computationally simpler and faster),
although the LoS direction is still needed to derive the perpendicular component of the field.
This is the approach used in this paper.\footnote{However, 
we have also implemented the LoS integration method and confirmed that the calculated total synchrotron specific luminosity 
is the same for both methods, and that synchrotron self-absorption and free-free self-absorption 
are negligible over the frequency range considered in this study.
This implementation will be discussed in the next paper in this series.}

We have verified that the luminosity statistics are converged with respect to the grid resolution.
We did this by rerunning \textsc{magnetizer} with twice and half the 
resolution used throughout this study (81 grid points in $r$ and larger when the disc radius is increasing),
finding negligible difference in the results.

\subsubsection{Synchrotron emissivity}
The number density of cosmic ray electrons with energies between $E$ and $E+\di E$ is given by 
\begin{equation}\label{eq:4}
  N(E)\,\di E = K_E E^{-s}\,\di E,
\end{equation}
where $K_E$ is a normalization factor and $s$ is the energy spectral index.
The synchrotron emissivity (the energy emitted by unit volume of the source 
per unit time and unit frequency interval within unit solid angle) in a homogeneous magnetic field $\bfB$
is given by \citep{Rybicki+Lightman79, Shukurov+Subramanian21}
\begin{equation}\label{eq:5}
  \epsilon = K_E \, a(s)\!\left( \frac{e^3}{m\elec c^2} \right)\!\left( \frac{3e}{4\uppi \, m\elec^3 \, c^5} \right)^{\frac{s-1}{2}}\!\!B^{\frac{s+1}{2}}_{\perp} \nu^{-\frac{s-1}{2}},
\end{equation}
where $\nu$ is the frequency of the radiation in the rest frame of the galaxy. 
We adopt $s=3$ but we also explore how the results are affected if $s$ is allowed to vary.
Here $B_{\perp}$ is the total magnetic field strength in the sky plane perpendicular to the LoS and 
\begin{equation}\label{eq:6}
  a(s) = \frac{\sqrt{3}}{4 \uppi \, (s+1)}\Gamma\left(\frac{3s-1}{12}\right)\Gamma\left(\frac{3s+19}{12}\right),
\end{equation}
where $\Gamma$ is the gamma function.
To obtain $K_E$, we assume energy equipartition between cosmic rays 
and the magnetic field \citep{Beck&Krause05},
\begin{equation}\label{eq:7}
  k\cosmray \int_{E_1}^{E_2} EN(E)\,\di E = \frac{B^2}{8 \uppi},
\end{equation}
where $k\cosmray$ is the ratio of the energy densities of the relativistic protons and electrons
and $B$ is the total magnetic field strength; below we use $k\cosmray=100$.
The
equipartition is in fact assumed only on scales greater than or equal to the model resolution,
which is typically $\gtrsim200\pc$ since \textsc{magnetizer} employs $81$ grid cells across the galactic radius.
Substituting equation~\eqref{eq:6} into equation~\eqref{eq:7} and integrating, we obtain
\begin{equation}\label{eq:8}
  K_E = \frac{B^2}{8 \uppi \, k\cosmray }\left(\frac{s-2}{E^{-s+2}_1-E^{-s+2}_2}\right),
\end{equation}
where $B^2 = \ol{B}^2 + b\rms^2$, $E_2 \to\infty$ and $E_1= 8\,\rm GeV$, 
the energy where the electron spectrum flattens in the solar neighbourhood \citep[e.g., section~10.3 of][and references therein]{Shukurov+Subramanian21}. 

\begin{figure}
\begin{center}
     \includegraphics [width=0.8\columnwidth]{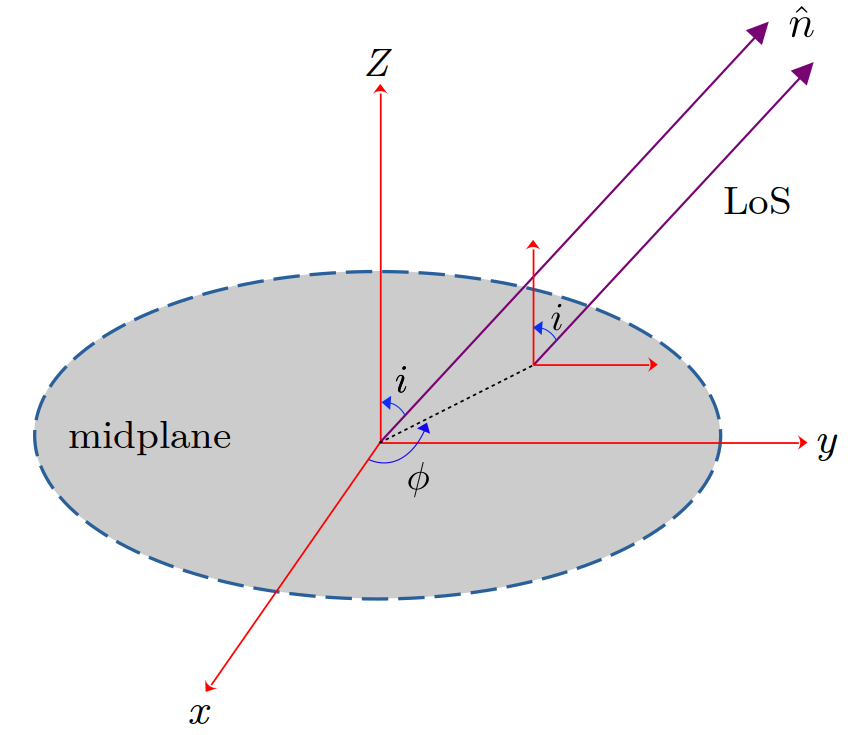}
     \caption{Schematic diagram of a galaxy disc 
     showing the line of sight (LoS) unit vectors $\hat{n}$ passing through the centre of the galaxy 
     and an arbitrary location $(r, \phi)$. 
     The galaxy is inclined at an angle $i$ with respect to the LoS. 
     The LoS directions are parallel to the $yZ$-plane in the galactic reference frame.}
     \label{fig:Los_midplane}
\end{center}
\end{figure}

\subsubsection{Computing $B_{\perp}$}
Let $i$ be the inclination angle ($0 \leq i \leq \uppi/2$) between the observer’s LoS 
and the normal to the galactic disc (the $Z$-axis; see Fig.~\ref{fig:Los_midplane}). 
The components of the unit vector $\bfnhat$ along any LoS passing 
through a point $(r, \phi, 0)$ in the disc plane are
\begin{equation}\label{eq:9}
  \bm{\hat{n}} \equiv (n_r,n_{\phi},n_Z) = (\cos \phi\sin i , -\sin \phi\sin i , \cos i).
\end{equation}
The perpendicular component of the mean magnetic field can be written as
\begin{equation}
  \meanv{B}_\perp = \meanv{B} - \bfnhat (\meanv{B} \cdot \bfnhat),
\end{equation}
where $\ol{B}_\parallel=\meanv{B} \cdot \bfnhat$ is the parallel component of the mean magnetic field, 
\begin{equation}\label{eq:11}
  \ol{B}_\parallel = \ol{B}_r(r)\cos\phi\sin i - \ol{B}_{\phi}(r)\sin \phi\sin i + \ol{B}_{Z}(r)\cos i.
\end{equation}
Thus, the square of the magnitude of $\meanv{B}_\perp$ is given by
\begin{equation}\label{eq:12}
  \ol{B}_{\perp}^2(r,\phi) = \ol{B}^2(r,\phi) - \ol{B}_\parallel^2(r,\phi),
\end{equation}
where $\ol{B}^2 (r,\phi)=\ol{B}_r^2(r,\phi) +\ol{B}_{\phi}^2(r,\phi)+\ol{B}_{Z}^2(r,\phi)$.  
If the turbulent magnetic field $\bm{b}$ is isotropic then $b_\perp^2 = \tfrac{2}{3}b\rms^2$ and we obtain
\begin{equation}\label{eq:13}
  B_{\perp}^2(r,\phi) = {\ol{B}_{\perp}}^2(r,\phi) + b_{\perp}^2 = {\ol{B}_{\perp}}^2(r,\phi) + \frac{2}{3}b\rms^2.
\end{equation}

\subsubsection{Total synchrotron specific luminosity}
At a given frequency $\nu$ (measured in the reference frame of the source),
the total synchrotron specific luminosity is given by
\begin{equation}\label{eq:21}
  \Lspec = 4\uppi \int_V \epsilon(\bm{r}) \,\di^3\bm{r} = 4\uppi \int_V \epsilon(r,\phi,Z)\,r\,\di r\, \di\phi\, \di Z.
\end{equation}
Following the assumption widely made in the interpretation of observations, we assume that the emission is isotropic. Although this is not true in the case of synchrotron emission, such an assumption can be acceptable when applied to a large sample of randomly oriented galaxies.

The integration in equation~\eqref{eq:21} is carried out over the volume of the disc only, excluding the (gaseous) galactic halo.
Synchrotron emission from galaxy haloes can be significant (albeit not dominant) in nearby galaxies \citep{Krause19,Beck+19}, 
but modelling it would require a more careful consideration of the galactic structure, 
including the effects of galactic winds and fountain flows, which are not included in our model.
We thus choose to leave this for future work.

\subsection{Selection criteria and inclination distribution}\label{sec:selection}
To identify star-forming, disc galaxies, we use the bulge-to-total stellar mass ratio $B/T$.  
Its median values for late-type spiral galaxies (Sa–-Sd) are observed to be typically below $0.4$, 
whereas early-type (E--S0) galaxies generally have $B/T \geq 0.6$ 
(see Appendix~\ref{app:B_T_ratio} for details). 
Galaxies with higher bulge-to-total ratios are more likely to host radio-loud AGN, 
reflecting the correlation between bulge mass and central black hole mass (Section 4.2.4 of \citetalias{Lacey+16} and reference therein).
Therefore, we select galaxies with $B/T \leq 0.4$, 
which helps to minimize the AGN contamination and isolate star formation–dominated radio emission.
We note that 
\citetalias{Jose+24} do
not use this selection criterion.
We apply this selection in all of our models, except for model~\texttt{J24}.

We assume that the galactic inclination angle has the probability distribution 
proportional to $\sin i$ with $0\leq i\leq \pi/2$, corresponding to the uniform distribution of orientations over a sphere. 
We also tried the uniform probability distribution of $i$ and a fit to the distribution of the inclination angles 
of the galaxies where it is known. 
These variations in the inclination distribution have a negligible effect on the results as long as the small-scale field dominates over the large-scale field. 
\begin{table*}
\centering
\footnotesize
\caption{
Summary of the models considered in this work. The columns list the model name, the magnetic field component(s) used to compute $\Lspec$, the ISM gas phases included, the cosmic star formation rate density (CSFRD) prescription, the turbulent speed model parameters ($v_0$ and $c$), the parameter $f_b$, the large-scale dynamo parameter $R_{\kappa}$, the dynamo type, the magnetic field scale height, and the interpretation of the no-$Z$ large-scale magnetic field.
}
\label{tab:runs}
\begin{tabular}{lcccccccccc}
\hline
\hline
Model &
Magnetic field &
ISM gas &
CSFRD &
$v_0$ &
$c$ &
$f_b$ &
$R_{\kappa}$ &
Dynamo &
Scale height &
Field interpretation \\
&
&
&
&
(km s$^{-1}$)
&
&
&
&
type
&
&
\\
\hline
\texttt{Fiducial} &
$\mathbf{\bar{B}}+\mathbf{b}$ &
All &
L16 &
15 &
0.3 &
0.8 &
0.3 &
$\alpha\Omega$ &
$2h_{\rm disk}$ &
Vertical average \\

\texttt{LS} &
$\mathbf{\bar{B}}$ &
All &
L16 &
15 &
0.3 &
0.8 &
0.3 &
$\alpha\Omega$ &
$2h_{\rm disk}$ &
Vertical average \\

\texttt{SS} &
$\mathbf{b}$ &
All &
L16 &
15 &
0.3 &
0.8 &
0.3 &
$\alpha\Omega$ &
$2h_{\rm disk}$ &
Not applicable \\

\texttt{J24} &
$\mathbf{\bar{B}}+\mathbf{b}$ &
Diffuse &
HB06 &
25 &
0.5 &
1.0 &
1.5 &
$\alpha^2\Omega$ &
$h_{\rm disk}$ &
Midplane \\
\hline
\end{tabular}
\end{table*}
\subsubsection{Star formation rate}\label{sec:SFR_threshold}
In \textsc{galform}, the star formation rate ($\SFR$) is assumed to be proportional to the molecular gas mass 
of galaxies, 
following the empirical relation of \citet{Blitz&Roso2006} (see \citetalias{Lacey+16} and \citealt{Lagos+13} for details). 
The model predicts a wide range of star formation rates, 
including distinct active (star-forming, gas-rich) and passive (quiescent, gas-poor) sequences, 
extending down to $\SFR \simeq 10^{-6}\Msunyr$.
However, at lower redshifts, \textsc{galform} tends to produce an extensive passive sequence, 
with galaxies exhibiting SFRs several orders of magnitude below the limits of current observational samples 
($\SFR \lesssim 10^{-3}\Msunyr$; \citetalias{Lacey+16}). 
Therefore, we consider only galaxies with $\SFR \geq 10^{-3}\Msunyr$, unless mentioned otherwise.
The \citetalias{Jose+24} model of \textsc{magnetizer} modifies the SFR obtained from \textsc{galform} 
by multiplying it by a redshift-dependent factor to bring the SFR density (SFRD, which is the SFR per unit co-moving volume)
into agreement with observations of \citet{Hopkins+Beacom06}.
The modified SFR is then used in equation~\eqref{eq: sfr-vt} to derive the turbulent speed.
This modification makes the model less consistent internally but more consistent with the observational SFR data. 
Therefore, for all of our models other than \texttt{J24} (see Section~\ref{sec:adjustable_parameters} for their description) 
we use the original \textsc{galform} output for the SFR. 

For computing the turbulent speed, we use equation~\eqref{eq: sfr-vt} with the disc SFR. 
However, the combined rate of quiescent star formation in the disc and bursty star formation \citepalias{Lacey+16} which can occur elsewhere is used to present our results since the total SFR is used when presenting observational data. 

\subsubsection{Stellar mass threshold}
\textsc{galform} models galaxies with stellar masses $M_\star$ between $10^6$ and $10^{12}\Msun$. 
However, most of the observed spiral galaxies considered in this work have $M_\star\gtrsim 10^9\Msun$, 
so we restrict our analysis to galaxies with $M_\star\geq 10^9\Msun$ unless stated otherwise. 
We discuss the sensitivity of our results to the stellar mass threshold in Section~\ref{sec:Mstar_dependence}.

\subsection{Adjustable parameters of the model}\label{sec:adjustable_parameters}
\label{sec:models}
There are two adjustable parameters of the model, 
viz.~$f_b$ and $R_\kappa$. Our assumed parameter values (Table~\ref{tab:runs}) 
are broadly consistent with those found using theoretical or computational models  
(e.g.~\citealt{Kim+Ostriker15b} and \citealt{Gent+24} for $f_b$ 
and \citealt{Mitra+10} and \citealt{Gopalakrishnan+Subramanian23} for $R_\kappa$).

The parameter $f_b$ affects the general level of the synchrotron intensity
but has a negligible effect on the power law exponents (i.e., the slopes of the log--log plots) 
of the $\Lspec$--$\SFR$ and $\Lspec$--$V\rot$ relations.
The parameter $R_\kappa$ does not strongly affect the results as long as the large-scale field 
is weaker than the small-scale field (a condition that is usually met).

The models considered are presented in Table~\ref{tab:runs}.
We focus our discussion on the fiducial (best-fit) model, named \texttt{Fiducial}.
In the interest of continuity and consistency with previous work, \citetalias{Jose+24}, 
we also present results for the model from that paper (model~\texttt{J24}). 
We adopt $f_b=0.8$ and $R_\kappa=0.3$ in all simulations other than in \texttt{J24}, where $f_b=1$ and $R_\kappa=1.5$.
We also considered models with only the large-scale magnetic field included (model~\texttt{LS}) 
and only the small-scale field included (model~\texttt{SS}),
to better understand the relative contributions of these two components. 

For \texttt{Fiducial}, we also adjust the parameters $c$ and $v_0$ of equation~\eqref{eq: sfr-vt} to produce a fit that is close to the median of the $v_{\rm t}\text{--}\rm SFR$ data and produces results that agree with observations (see Appendix~\ref{app:v_turb-SFR}). The results are insensitive to $v_0$ but the choice of the parameter $c$ moderately affects the scaling relation exponents for galaxies with $\rm{SFR} \geq 1\,M_{\odot}\,\rm{yr}^{-1}$.

\begin{deluxetable*}{ccccccccccc}
\tabletypesize{\footnotesize}
\setlength{\tabcolsep}{3pt}
\tablecaption{Basic properties and synchrotron specific luminosities at different frequencies of the observed galaxy sample. 
\label{tab:data}}

\tablehead{
\colhead{\makecell{Galaxy \\ name}} &
\colhead{\makecell{Hubble \\ type}} &
\colhead{\makecell{$d$\\$[\rm Mpc]$}} &
\colhead{\makecell{$\log {M_\star}$ \\ $[\rm M_{\odot}]$}}
&
\colhead{\makecell{$\SFR$\\$[\Msunyr]$} }&
\colhead{\makecell{$V_{\rm rot}$\\$[\rm km\,s^{-1}]$}} &
\colhead{$\log\!\tfrac{L_{10.7}}{[\mathrm{W\,Hz^{-1}}]}$} &
\colhead{$\log\!\tfrac{L_{8.4}}{[\mathrm{W\,Hz^{-1}}]}$} &
\colhead{$\log\!\frac{L_{4.8}}{[\mathrm{W\,Hz^{-1}}]}$} &
\colhead{$\log\!\tfrac{L_{1.4}}{[\mathrm{W\,Hz^{-1}}]}$ }&
\colhead{$\log\!\tfrac{L_{0.144}}{[\mathrm{W\,Hz^{-1}}]}$}
}
\startdata
DDO 53 & Im  & 3.61 & $6.25^{1}$ & $0.006^{\rm{b}}$ & $29^{\rm{c}}$ & ... & ... &$17.98 \pm 0.16^{\rm{b}}$ & ... & ... \\
DDO 154 & IBm & 4.30 & $6.63^{1}$ & $0.002^{\rm{b}}$ & $50^{\rm{c}}$ & ... & ... & $<17.89^{\rm{b}}$ & $<18.42^{\rm{b}}$ & ... \\
DDO 165 & Im  & 4.57 & $7.04^{1}$ & $0.002^{\rm{b}}$ & $20^{\rm{c}}$ & ... & ... &$<17.92^{\rm{b}}$ & $<18.46^{\rm{b}}$ & ... \\
HoI                                                    & IABm                                                   & 3.90                                                         & $6.87^{1}$                                & $0.004^{\rm{b}}$                                                                      & $53^{\rm{c}}$             & ... & ...                                                             & $18.19 \pm 0.27^{\rm{b}}$                                                                  & $<18.33^{\rm{b}}$                                                                        & ...                                                                                \\
IC 10                                                  & Irr                                                    & 0.66                                                         & $8.69^{19}$                               & $0.06^{\rm{a}}$                                                                       & $47\pm 5^{\rm{a}}$      & ... & ...                                                               & $18.91 \pm 0.08^{\rm{a}}$                                                                 & ...                                                                                & $19.63 \pm 0.04^{\rm{e}}$                                                                                \\
\enddata
\end{deluxetable*}
\newpage
\addtocounter{table}{-1}
\begin{deluxetable*}{ccccccccccc}
\tabletypesize{\footnotesize}
\setlength{\tabcolsep}{3pt}
\tablecaption{(continued)}
\tablehead{
\colhead{\makecell{Galaxy \\ name}} &
\colhead{\makecell{Hubble \\ type}} &
\colhead{\makecell{$d$\\$[\rm Mpc]$}} &
\colhead{\makecell{$\log {M_\star}$ \\ $[\rm M_{\odot}]$}}
&
\colhead{\makecell{$\SFR$\\$[\Msunyr]$} }&
\colhead{\makecell{$V_{\rm rot}$\\$[\rm km\,s^{-1}]$}} &
\colhead{$\log\!\tfrac{L_{10.7}}{[\mathrm{W\,Hz^{-1}}]}$} &
\colhead{$\log\!\tfrac{L_{8.4}}{[\mathrm{W\,Hz^{-1}}]}$} &
\colhead{$\log\!\frac{L_{4.8}}{[\mathrm{W\,Hz^{-1}}]}$} &
\colhead{$\log\!\tfrac{L_{1.4}}{[\mathrm{W\,Hz^{-1}}]}$ }&
\colhead{$\log\!\tfrac{L_{0.144}}{[\mathrm{W\,Hz^{-1}}]}$}
}
\startdata
NGC 4565                                               & SAb                                                    & 13.1                                                         & $10.90^{5}$                               & $0.36^{\rm{a}}$                                                                       & $255\pm 12^{\rm{a}}$                  &...&...                                                 & $21.94 \pm 0.05^{\rm{a}}$                                                                     & ...                                                                                & $22.22 \pm 0.04^{\rm{e}}$                                                                               \\
NGC 4569                                                & SABab                                                  & 9.86                                                         & $10.00^{1}$                               & $0.29^{\rm{b}}$                                                                       & ...                                  &$20.29 \pm 0.15^{\rm{b}}$    &$20.42 \pm 0.19^{\rm{b}}$                                         & $20.70 \pm 0.20^{\rm{b}}$                                                                & $21.51^{\rm d}$                                                                                & ...                                                                                \\
\enddata
\tablecomments{
The Hubble type and distance $d$ are taken from the \href{https://ned.ipac.caltech.edu/}{NASA Extragalactic Database}.  
a: \citet{Tabatabaei+16},
b: \citet{Tabatabaei+17},
c: \citet{Heesen+22},
d: \citet{Liu+15},
e: \citet{Heesen+26},
1: \citet{Skibba+2011},
2: \citet{Wang+2021},
3: \citet{Lucero+15},
4: \citet{BaesViaene2016},
5: \citet{Das+23},
6: \citet{Just+2006},
7: \citet{Harmsen+2017},
8: \citet{Stuber+2023},
9: \citet{Rozas+2008},
10: \citet{Foord+2017},
11: \citet{Groves+2023},
12: \citet{Wood+2015},
13: \citet{Muraoka+2019},
15: \citet{Sick+2015},
16: \citet{Corbelli+2014},
17: \citet{Erkal+2019},
18: \citet{Besla+2012},
19: \citet{Yin+2010},
20: \citet{Belland+2020},
21: \citet{Zheng+2020},
22: \citet{Zheng+2022},
23: \citet{Wang+2016},
24: \citet{Gilhuly+2022},
25: \citet{Leroy+08},
26: \citet{Vollmer+2016},
27: \citet{Ma+2023}.
}
\end{deluxetable*}
\section{Observational data}\label{sec:data}
\subsection{Low-redshift galaxies}
To weaken observational selection effects 
and to obtain a statistically meaningful sample of galaxies across a range of observational frequencies, 
we combine the radio luminosity data from \cite{Liu+15}, \citetalias{Tabatabaei+16}, 
\citetalias{Tabatabaei+17}, \citetalias{Heesen+22}, and  \citet{Heesen+26} and present them in Table~\ref{tab:data}.
This sample 
includes $21$ edge-on ($i>75^{\circ}$) spiral galaxies out of $35$ from the CHANG-ES survey \citep{Irwin+12,Irwin+24}.
While this survey acquired data for L and C-bands \citep{Wiegert+15}, we choose not to include these data as they are unlikely to significantly change our overall results.
At the frequency $1.4\GHz$, we also consider a large set of nearby ($z\lesssim 0.02$) spiral galaxies (Sa--Sd) 
from \cite{Bell+03}, not listed in Table~\ref{tab:data}. 
For these, we have excluded the galaxies for which the data are already listed in Table~\ref{tab:data}.
To convert 
flux density $S_\nu$ to luminosity, we use $\Lspec = 4\uppi d^2 S_\nu$,
where $d$ is the distance to the galaxy. 

Most of the galaxies in the sample are classified as spirals; 
those few that are classified as irregulars, ellipticals and lenticulars 
are not as suitable for comparison with our modelled galaxies 
and thus are distinguished by special symbols in the $\Lspec$--$\SFR$ plots presented below,
and are not included in the data analysis.\footnote{We do not include the irregular, elliptical and lenticular galaxies in the $\Lspec$ versus $V\rot$ plots as they have very small rotation speeds.} 

SFR data presented in Table \ref{tab:data} 
are taken from \citetalias{Tabatabaei+16}, \citetalias{Heesen+22} and \citetalias{Tabatabaei+17}.
Three galaxies, IC~342, NGC~6469 and NGC~4536, 
are common to \citetalias{Tabatabaei+16} and \citetalias{Tabatabaei+17}; 
we opt to use \citetalias{Tabatabaei+17} for the luminosity and SFR data of these galaxies. 
For the data of \cite{Bell+03}, we compute the SFR from the total infrared (IR) luminosities 
using equation~(5) of that paper.

We subtract the thermal fraction from the observed radio emission to obtain the synchrotron luminosity. 
For the sample of \citetalias{Tabatabaei+17}, 
the thermal fraction at $1.4\GHz$ and $4.8\GHz$ is given by the authors for most of the galaxies.
Where the thermal fraction has not been published, including the galaxies of \cite{Bell+03},
we use the averages of the thermal fractions at those frequencies from \citetalias{Tabatabaei+17}. 
At $10.7\GHz$ and $8.4\GHz$, the thermal fractions for an optically thin medium are obtained as\footnote{The radio 
continuum spectrum can be written as 
$L_\nu^\mathrm{tot} = L_\nu^\mathrm{th} + \Lspec = A_1\,\nu^{-0.1} + A_2\,\nu^{-\alpha_{\rm nt}}$, 
where $\alpha_{\rm nt}$ is the nonthermal spectral index and $A$ and $B$ are constant scaling factors. 
The thermal term in this expression represents the Planck function for 
an optically thin ionized gas \citepalias{Tabatabaei+17}.}
\begin{equation}
   f_{\rm \nu}^{\rm th} = f_{\nu_0}^{\rm th}\left(\frac{L_{\nu_0}}{L_\nu}\right)\left(\frac{\nu}{\nu_0}\right)^{-\alpha_{\rm th}},
\end{equation}
where   $f_{\nu_0}^{\rm th}$ and $L_{\nu_0}$ are known \citepalias{Tabatabaei+17}, 
$L_{\nu}$ is the total specific luminosity at frequency $\nu$ including both the thermal and non-thermal contributions,
$\nu_0$ can be $4.8\GHz$ or $1.4\GHz$ and $\alpha_{\rm th}$ is the spectral index of the thermal emission;
for an optically thin ionized medium $\alpha_{\rm th} \approx0.1$.
At $144\MHz$, however, 
we use the total specific luminosities without correction since the thermal contribution is negligible. 

Rotation speed data for the galaxies from \citetalias{Tabatabaei+16} and \citetalias{Heesen+22} 
are provided in those references.
For the galaxies studied in \citetalias{Tabatabaei+17}, 
we adopt the rotation speeds from \citetalias{Heesen+22}, when available. 
The galaxy NGC~5474 is a dwarf satellite with a much lower rotation speed than typical spiral galaxies, 
despite exhibiting a high specific star formation rate $\text{sSFR} > 10^{-10.4}\yr^{-1}$).
\citet{Bellazzini+20} classify NGC~5474 as a peculiar spiral galaxy.
We choose to exclude this galaxy when investigating the relationship between $\Lspec$ and $V_{\rm rot}$.

\subsection{MIGHTEE-COSMOS galaxies}\label{MCD}
We also use the MIGHTEE–COSMOS survey data presented by \citet{An+21} at 1.3\,GHz (observer frame). 
These authors combined the 1.3\,GHz MIGHTEE survey data with the `super-deblended' catalogue of \cite{Jin+18}. 
From this dataset, we select only SFGs within the redshift range $z = 0$--$3$ 
by excluding AGNs and red quiescent galaxies (for details, see \citealt{An+21}).
We derive the thermal fraction $f^{\rm th}_{\nu_{\rm rest}}$ at the rest-frame frequency $\nu_{\rm rest} = 1.3(1+z)\GHz$ 
using the average thermal fraction $f^{\rm th}_{4.8}$ at $4.8\GHz$
from the \citetalias{Tabatabaei+17} galaxy sample, rescaled as (see Appendix~\ref{app:thermal_frac})
\begin{equation}
    f^{\rm th}_{\nu_{\rm rest}} 
    = \left[1+ \left(\frac{1}{f_{4.8}^{\rm th}}-1\right)
    \left(\frac{\nu_{\rm rest}}{4.8\GHz}\right)^{-(\alpha_{\rm nt}-\alpha_{\rm th})}\right]^{-1},
\end{equation}
where $\alpha_{\rm nt}$ and $\alpha_{\rm th}$ are the nonthermal and thermal spectral indices, respectively. 
In our model, $\alpha_{\rm th}=0.1$  
and $\alpha_{\rm nt}=1$, which follows from $s=3$. 
For consistency, we continue to use $\alpha_{\rm nt}=1$ at high redshift,
though there is some evidence that $\alpha_\mathrm{nt}$ decreases with $\sSFR$:
for the COSMOS survey, it is found to vary in the range $0.2\lesssim\alpha_\mathrm{nt}\lesssim1.3$
with mean value $\langle\alpha_\mathrm{nt}\rangle\sim0.75$ \citep{Tabatabaei+25}.
However, we verify that this choice does not affect our conclusions. 
For more detailed estimates of the thermal fraction based on spectral energy distribution (SED) modeling, 
we refer to \citetalias{Tabatabaei+17} and \citet{Tabatabaei+25}.

We subtract the thermal component from the observed $1.3\GHz$ flux densities. 
The rest-frame nonthermal luminosity, 
$L_{\nu_{\rm rest}}$, is then obtained as, 
\begin{equation}
    L_{\nu_{\rm rest}} = \frac{4\pi d^2}{1+z} S_{\nu_{\rm obs}},
\end{equation}
where $d$ is the luminosity distance (computed using flat $\Lambda\rm CDM$ cosmology; $H_0 = 70\rm \km\,s^{-1} \, Mpc^{-1}$, $\Omega_m =0.3$ , and $\Omega_{\Lambda}= 0.7$) and 
$S_{\nu_{\rm obs}}$ 
is the synchrotron flux density in the observer's frame.
The SFR (estimated from the total IR emission) and stellar mass data are also available in the same catalogue.

\section{Results}\label{sec:results}
\begin{table*}
\tabletypesize{\tiny}
\setlength{\tabcolsep}{2.5pt}
    \centering
    \footnotesize
    \caption{The Spearman rank correlation coefficient $r_{\rm sp}$, the Pearson correlation coefficient $r_{\rm p}$ and power law index (slope) $\gamma$ for the $\Lspec$--$\SFR$ and $\Lspec$--$V\rot$ relations of the form of equation~\eqref{eq:fit_line}, as specified in the first row for the models considered in the text. We have only listed the normalization of the best-fitting line (equation~\ref{eq:fit_line}) for $\nu = 1.4\,$GHz, $L_{0,\,1.4}$ at $\rm SFR= 0.1\,M_{\odot}\,yr^{-1}$ and $100\,\rm km\,s^{-1}$ for $L_\nu\text{--}\rm SFR$ and $L_\nu\text{--}V\rot$, respectively. The values of normalization at another frequency $\nu$ can be calculated by multiplying $L_{0,\,1.4}$ by $1.4\,\mathrm{GHz}/\nu$.}
    \tabletypesize{\small}
    \label{tab:correlations}
    \begin{tabular*}{\textwidth}{@{\extracolsep{\fill}}ccccccccc}
\hline\hline
 & \multicolumn{4}{c}{$\Lspec$--$\SFR$} 
 & \multicolumn{4}{c}{$\Lspec$--$V_{\rm rot}$} \\
\cline{2-5}\cline{6-9}
 Model& \texttt{Fiducial} & \texttt{J24} & \texttt{LS} & \texttt{SS}
 & \texttt{Fiducial} & \texttt{J24} & \texttt{LS} & \texttt{SS} \\
\hline
$r_{\rm sp}$ &
0.81 & 0.82 & 0.57 & 0.91 &
0.53 & 0.85 & 0.25 & 0.71 \\
$r_{\rm p}$ &
0.90 & 0.84 & 0.62 & 0.95 &
0.54 & 0.84 & 0.23 & 0.71 \\
$\gamma$ &
1.00 & 0.65 & 0.81 & 1.08 &
3.43 & 2.84 & 2.72 & 3.56 \\
$\log(L_{0,1.4})$ &
20.09 & 20.27 & 19.12 & 19.71 &
20.14 & 20.19 & 19.21 & 19.79 \\
\hline
\end{tabular*}
\end{table*}
\begin{table*}
\centering
\footnotesize
\setlength{\tabcolsep}{2.5pt}
\caption{
The Pearson and Spearman correlation coefficients $r_\mathrm{p}$ and $r_\mathrm{sp}$, respectively,
and the best-fitting slope $\gamma$ and normalization $L_{0, \nu}$ for the fit $L_\nu = L_0 (x/x_0)^\gamma$,
where $x=\mathrm{SFR}$ or $V_\mathrm{rot}$ and $x_0=0.1\,M_\odot\,\mathrm{yr}^{-1}$ or $100\,\mathrm{km\,s^{-1}}$, respectively, computed using the observational data.
For the $L_\nu$--$V_{\rm rot}$ relation at $8.4\,\mathrm{GHz}$,
we do not compute the correlation coefficients and slope because the number of observational data points
is too small to get a statistically reliable estimate.
}
\label{tab:obs_data_corr}
\begin{tabular}{ccccccrcccc}
\hline\hline
 & \multicolumn{5}{c}{$L_\nu$--SFR}
 & & \multicolumn{4}{c}{$L_\nu$--$V_{\rm rot}$} \\
 \cline{2-6}\cline{8-11}
 & 10.7 GHz & 8.4 GHz & 4.8 GHz & 1.4 GHz & 144 MHz & & 10.7 GHz & 4.8 GHz & 1.4 GHz & 144 MHz \\
\hline
$r_{\rm sp}$ &
$0.68\pm0.19$ &
$0.85\pm0.27$ &
$0.82\pm0.14$ &
$0.86\pm0.08$&
$0.84\pm0.14$ & &
$0.48\pm0.24$ &
$0.49\pm0.19$ &
 $0.61\pm0.19$ &
$0.44\pm0.18$ \\
$r_{\rm p}$  &
$0.70\pm0.14$ &
$0.90\pm0.12$ &
$0.85\pm0.07$ &
$0.87\pm0.04$ &
$0.86\pm0.07$ & &
$0.54\pm0.20$ &
$0.63\pm0.15$ &
$0.64\pm0.15$ &
$0.54\pm0.16$ \\
$\gamma$     &
$0.97\pm0.19$ &
$1.10\pm0.15$ &
$1.08\pm0.09$ &
$1.17\pm0.10$ &
$1.24\pm0.11$ & &
$2.55\pm0.96$ &
$2.79\pm0.66$ &
$2.90\pm0.69$ &
$2.77\pm0.80$ \\
$\log(L_{0, \nu})$        &
$19.55\pm0.21$&
$19.47\pm0.19$&
$19.77\pm0.10$&
$19.96\pm 0.06$&
$20.74\pm0.12$& &
$19.83\pm0.30$&
$20.25\pm0.21$&
$20.74\pm0.22$&
$21.37\pm0.25$\\
\hline
\end{tabular}
\end{table*}
\begin{table*}
    \centering
    \footnotesize
    \caption{Comparison of the observational data-based correlation coefficients at $4.8\,$GHz found in \citetalias{Tabatabaei+16} with those obtained in this work using spiral galaxies alone from the combined observational data set.}
    \label{tab:T16_comp}
    \begin{tabular*}{\textwidth}{@{\extracolsep{\fill}} l ccc c ccc }
        \hline\hline
         & \multicolumn{3}{c}{$\Lspec$--$\SFR$} & & \multicolumn{3}{c}{$\Lspec$--$V_{\rm rot}$} \\
        \cline{2-4}\cline{6-8}
         & $r_{\rm sp}$ & $r_{\rm p}$ & $\gamma$ & & $r_{\rm sp}$ & $r_{\rm p}$ & $\gamma$ \\
        \hline
        T16                 & $0.91\pm0.05$ & $0.86\pm0.09$ & $1.22\pm0.12$ & & $0.72\pm0.01$ & $0.85\pm0.05$ & $4.00\pm0.30$ \\
        This work (spirals) & $0.82\pm0.14$ & $0.85\pm0.07$ & $1.08\pm0.09$ & & $0.49\pm0.19$ & $0.63\pm0.15$ & $2.79\pm0.66$ \\
        \hline
    \end{tabular*}
\end{table*}
We focus on the correlations between the total synchrotron specific luminosity $\Lspec$ and $\SFR$, 
and $\Lspec$ and the rotation speed in the flat part of the rotation curve $V\rot$, and compare our results with observational data for nearby galaxies and also for higher redshifts wherever observational data are available.
For statistical analysis, we use a randomly selected sample of $2\times10^5$ simulated galaxies,
tracing their evolution from $z=6$ to $0$.
However, convergence is typically achieved with as few as about $3 \times 10^4$ galaxies. 

Results of the correlation analysis for the different models and observational data 
are presented in Tables~\ref{tab:correlations} and \ref{tab:obs_data_corr}, respectively.
We calculate both the Pearson correlation coefficient $r\Pe$ (a measure of linear correlation)
and Spearman's rank coefficient $r\Sp$ (a measure of monotonic relationship).
We also include the parameters of the best-fitting linear model $\gamma$ and $L_{0, \nu}$, 
corresponding to the power law
\begin{equation}\label{eq:fit_line}
  \Lspec = L_{0,\nu}\left(\frac{x}{x_0}\right)^\gamma,
\end{equation}
where $x$ and $x_0$ are either $\SFR$ and $0.1\Msunyr$ or $V\rot$ and $100\kms$, respectively.
In our model, $r\Pe$, $r\Sp$ and $\gamma$ are independent of the frequency 
since we assume that the spectral index of the cosmic ray electron spectrum is constant, viz.~$s=3$.
From equations~\eqref{eq:5} and \eqref{eq:21}, 
the synchrotron luminosity scales as $\nu^{-1}$ for $s=3$,
so the normalization $L_{0, \nu}$ varies with frequency.
Therefore, for comparison with the observational data, we present separate plots for different frequencies.
\subsection{Comparison of empirical correlations with T16 results}\label{sec: T16_comp}
While \citetalias{Tabatabaei+16} focused only on isolated galaxies, 
the combined data set also includes galaxies in cluster environments, as well as both satellite and central galaxies. 
The correlation coefficients and slopes for the \citetalias{Tabatabaei+16} sample 
and for spiral galaxies from  the combined sample at $4.8\GHz$ are presented in Table~\ref{tab:T16_comp}. 
For the $\Lspec$--$\SFR$ correlation, the empirical correlations from the combined data set agree well with those of \citetalias{Tabatabaei+16}.
However, the $\Lspec$--$V_{\rm rot}$ correlations are weaker and the slope is lower for the combined data set.
This difference may be partly caused by the inclusion of irregular 
galaxies in the sample of \citetalias{Tabatabaei+16}.


\subsection{Correlation between $\Lspec$ and $\SFR$}\label{sec:SFR}
\begin{figure*}[ht!]
    \centering
    \subfigure{
        \includegraphics[width=0.485\textwidth]{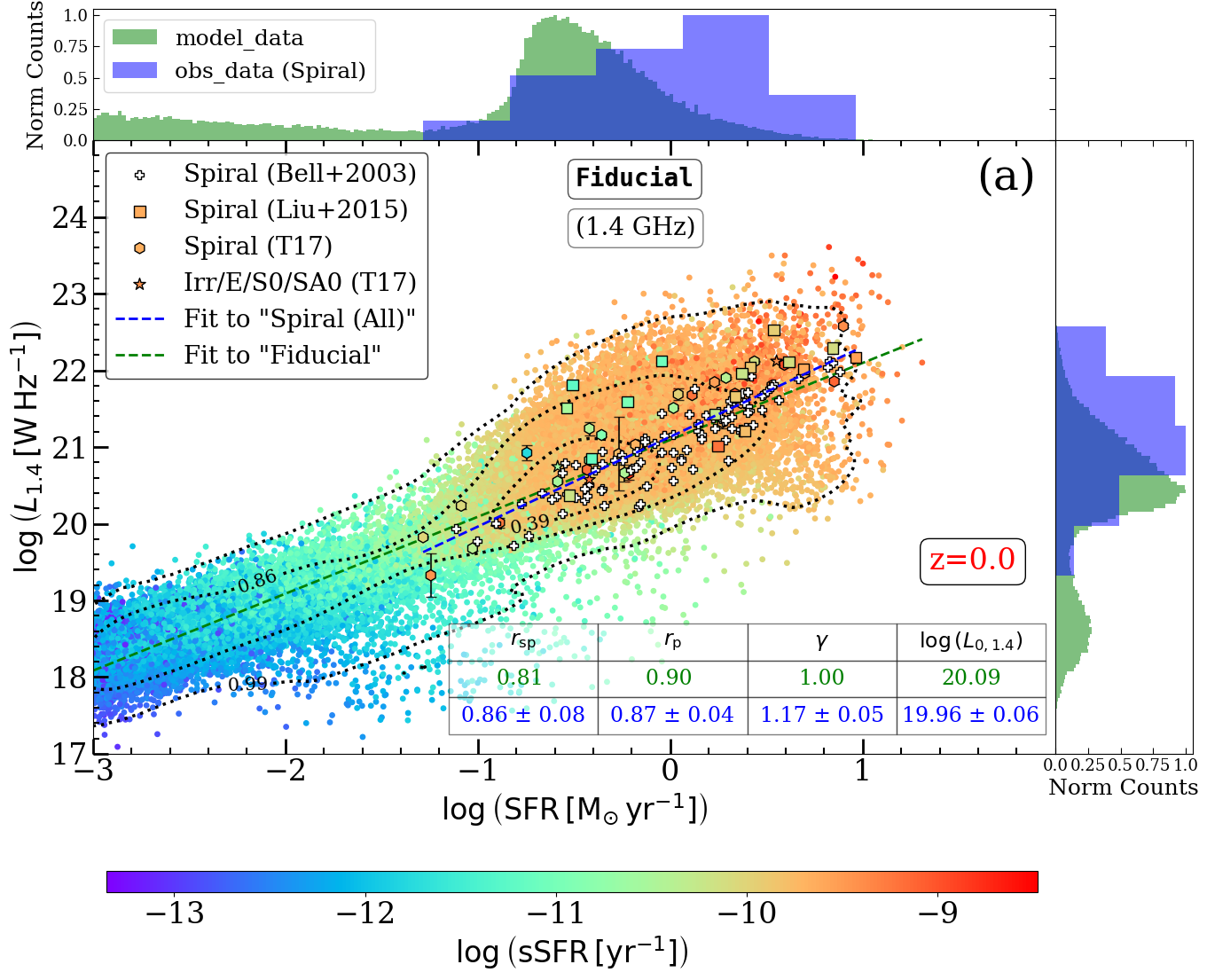}
        \label{fig: SI vs SFR 4.8 b}
    }
    \subfigure{
        \includegraphics[width=0.485\textwidth]{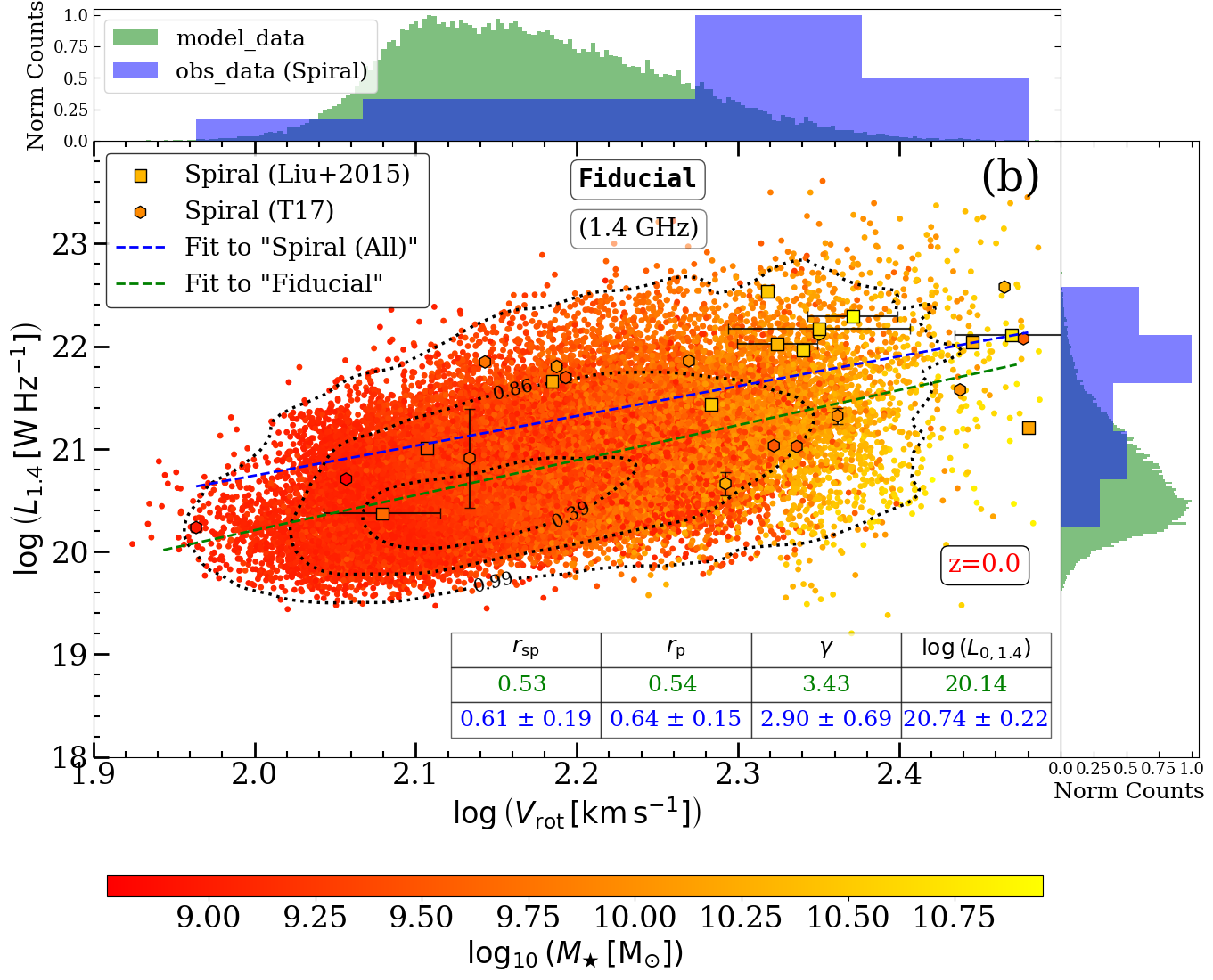}
        \label{fig:Lum_I_Vrot_4.8 b}
    }
    \caption{
    Scatter plots comparing the rest-frame $1.4\GHz$ total synchrotron luminosity, $\Lspec$, with the star formation rate, $\rm SFR$ (panel~a), and the rotational velocity in the flat part of the rotation curve, $V\rot$, for actively star-forming galaxies with specific star formation rates $\sSFR>10^{-10.4}\yr^{-1}$ (panel~b) at redshift $z = 0$. 
    The small colored circles in both plots represent the simulated galaxies from the \texttt{Fiducial} model, while the different symbols with error bars correspond to the observational data.
    The color coding indicates the specific star formation rate ($\rm sSFR$) of the galaxies in panel~(a) and the stellar mass ($M_{\star}$) in panel~(b). 
    The histograms along the top axes show the probability distributions of the $\SFR$ (panel~a), and $V\rot$ (panel~b), whereas the histograms along the right axes show the probability distribution of $\Lspec$. In all histograms, blue represents the observational data and green the simulated data.
    The observational data for spiral galaxies are represented by plus symbols for \citet{Bell+03}, squares for \citet{Liu+15}, and hexagons for \citetalias{Tabatabaei+17}. The
    star symbols represents the irregular/elliptical/lenticular galaxies from \citetalias{Tabatabaei+17}. 
    The green and blue dashed lines represent power-law fits to the medians of the simulated and observed spiral galaxies, 
    respectively. 
    Contours represent the $39\%$, $86\%$ and $99\%$ confidence levels of the 2D kernel density estimate. 
    The inset tables summarize the Spearman ($r\Sp$) and Pearson ($r\Pe$) correlation coefficients, together with the best-fitting power-law slope ($\gamma$) and normalization ($L_{0,\, 1.4}$), evaluated at $\SFR=0.01 \Msunyr$ (for panel~a) and at $V\rot=100\kms$ (for panel~b) using the fitting function given by equation~\eqref{eq:fit_line},
    with green entries corresponding to the simulated sample and blue to the observational sample. These values are also listed in Table~\ref{tab:correlations}.
    }
    \label{fig: SI vs SFR 4.8}
\end{figure*}
\begin{figure}
    \centering
    \includegraphics[width=\columnwidth]{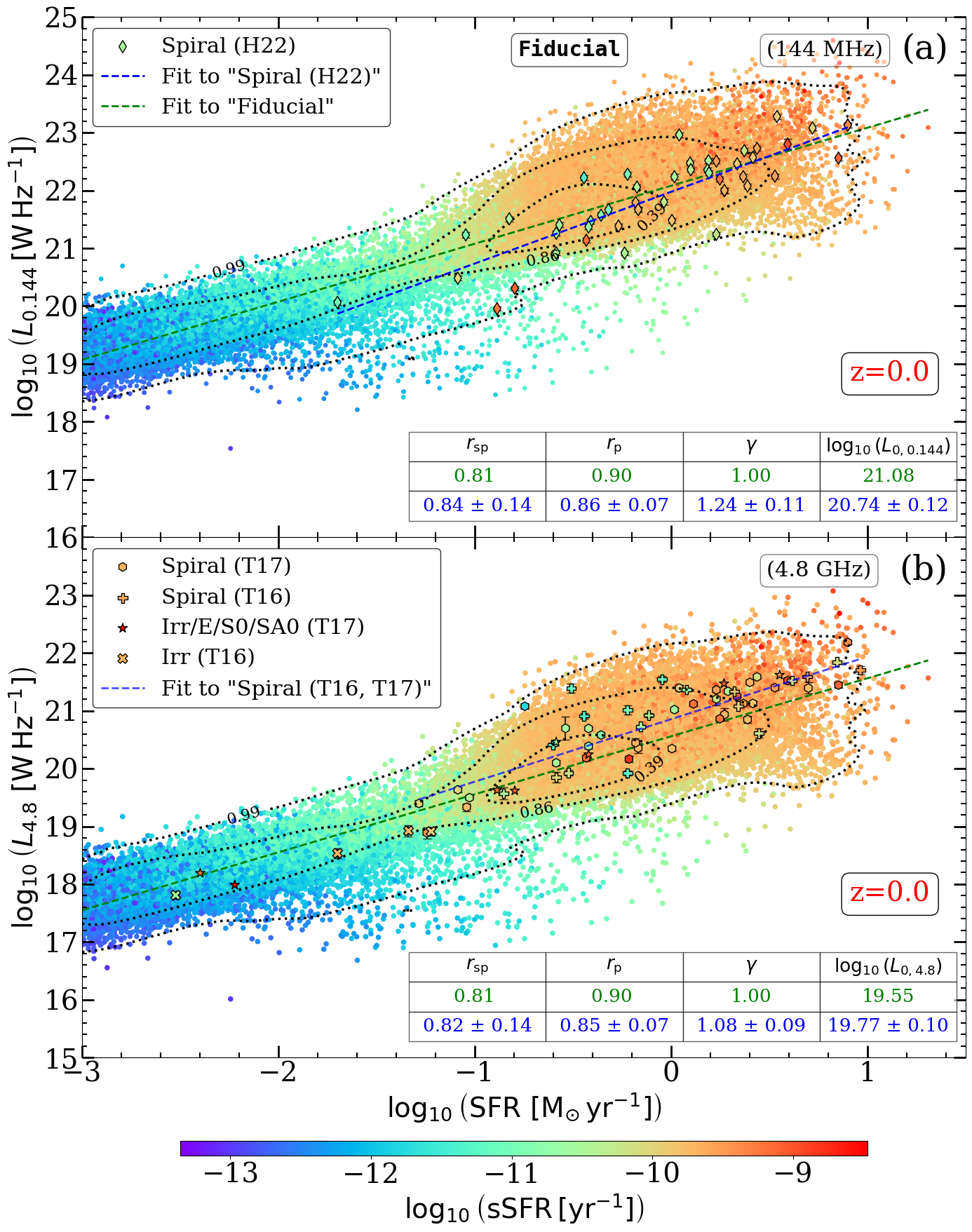}
    \caption{As Fig.~\ref{fig: SI vs SFR 4.8 b}, but for the \texttt{Fiducial} model at two other frequencies: 
    $144\rm \,MHz$ (panel~a) and $4.8 \GHz$ (panel~b). 
    In panel~(a), spiral galaxies from \citetalias{Heesen+22} are shown as diamonds.
    In panel~(b), 
    plus and hexagon symbols 
    represent the spiral galaxies and square and star symbols represent the irregular, elliptical and lenticular galaxies from \citet{Tabatabaei+16} (T16) and \citet{Tabatabaei+17} (T17), respectively.} 
    
    \label{fig: SI vs SFR 1.4 144}
\end{figure}

Figure~\ref{fig: SI vs SFR 4.8 b} presents a log-log plot of the total synchrotron luminosity versus the star formation rate  
at redshift $z=0$ and $1.4\GHz$ for our \texttt{Fiducial} model. 
Similar plots are presented in Figs.~\ref{fig: SI vs SFR 1.4 144}a--\ref{fig: SI vs SFR 1.4 144}b 
for \texttt{Fiducial}  
at $144\MHz$ and $4.8\GHz$, respectively. 
Because of the limited observational data available at $8.4\GHz$ and $10.7\GHz$, 
the corresponding plots are not shown. 
However, 
the correlation coefficients and best-fit parameters for these frequencies are provided in Table~\ref{tab:obs_data_corr}.
For determining the fit and calculating the correlation coefficients for the observational data, 
we choose to exclude irregular galaxies, as they are not modelled by \textsc{magnetizer}. 
The data for irregular galaxies (excluding those with luminosity upper limits) are shown using different symbols (see the plot legend). 

Both $r\Pe$ and $r\Sp$ are high for our 
\texttt{Fiducial} model,
suggesting a good agreement with observations. 
Comparing the color coding of the model points and observational data, 
it can be seen that 
our model is in good agreement with the observed $\sSFR$ trend.
However, while a few low-$\sSFR$ (quiescent) galaxies 
appear at the upper end of this trend in the data (green data points at relatively high $\SFR$),
such cases are far rarer in the simulations 
(they exist at some level but are visually obscured by the larger population of actively star-forming galaxies, 
colored orange-red).

We also show the histograms of the distributions of the model galaxies (green) and observed spiral galaxies (blue) 
in SFR and luminosity. 
Our \texttt{Fiducial} model exhibits a bimodal distribution in $\SFR$ and in $\Lspec$. 
We get a sharp peak at high $\SFR$ ($\SFR \simeq 4\times10^{-2}\Msunyr$) 
and high $\Lspec$ ($\Lspec \simeq 5\times10^{20}\,(\nu/1.4\GHz)^{-1}\,{\rm W\,Hz^{-1}}$), 
and a shallower peak at low luminosity $\Lspec \simeq 6\times10^{18}\,(\nu/{\rm 1.4 \,GHz})^{-1}\,{\rm W\,Hz^{-1}}$. 
In contrast, the observational data are confined mainly to the high-$\SFR$, high-$\Lspec$ region, 
likely due to observational selection effects (see Section~\ref{sec:obs_selec_effects}).
Nevertheless, the shape and standard deviation of the distributions at the high-$\SFR$--high-$\Lspec$ end 
are in reasonably good agreement with observations. 
For continuation, we also plot this correlation for the \texttt{J24} model in Figure~\ref{fig: SI vs SFR 4.8 a} of Appendix~\ref{App: Alt_models}. 
\texttt{J24} gives a smaller slope for the fitted median 
and lower luminosities for actively star-forming galaxies as compared to the observations,
whereas \texttt{Fiducial} agrees better in
both the slope and the luminosity magnitude.

%
\begin{figure}
    \centering
    
    \includegraphics[width=\columnwidth]{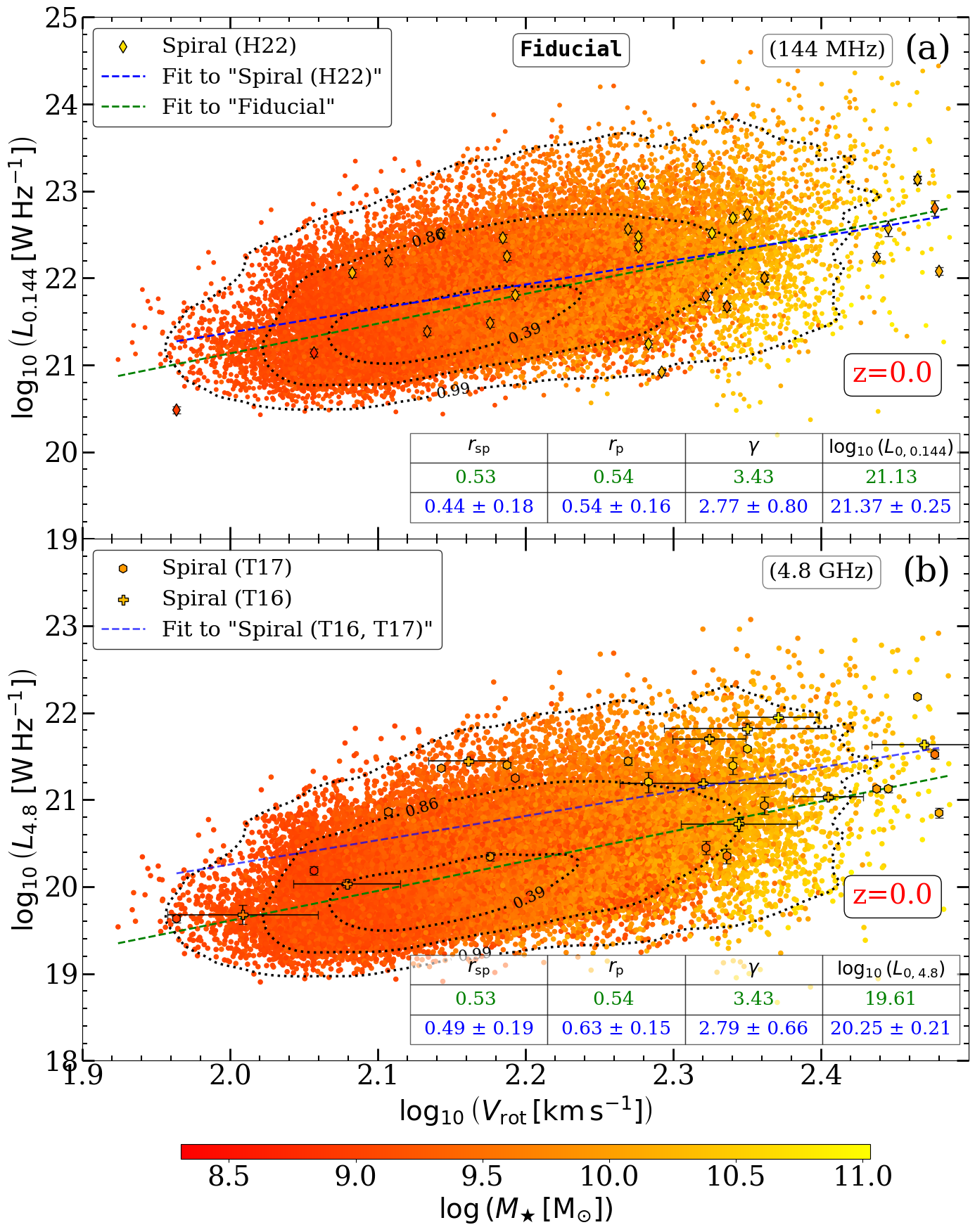}
    \caption{As Fig.~\ref{fig:Lum_I_Vrot_4.8 b}, for two different frequencies: 
    $144\MHz$ (panel~a), and $4.8\GHz$ (panel~b).
    In panel (a), the diamond symbols denote the spiral galaxies from \citetalias{Heesen+22}. In panel (b), the hexagon and plus symbols represent the spiral galaxies from \citetalias{Tabatabaei+16} and \citetalias{Tabatabaei+17}, respectively.}
    \label{fig:fig5}
\end{figure}
\begin{figure*}
\centering
    \includegraphics[width=0.90\textwidth]{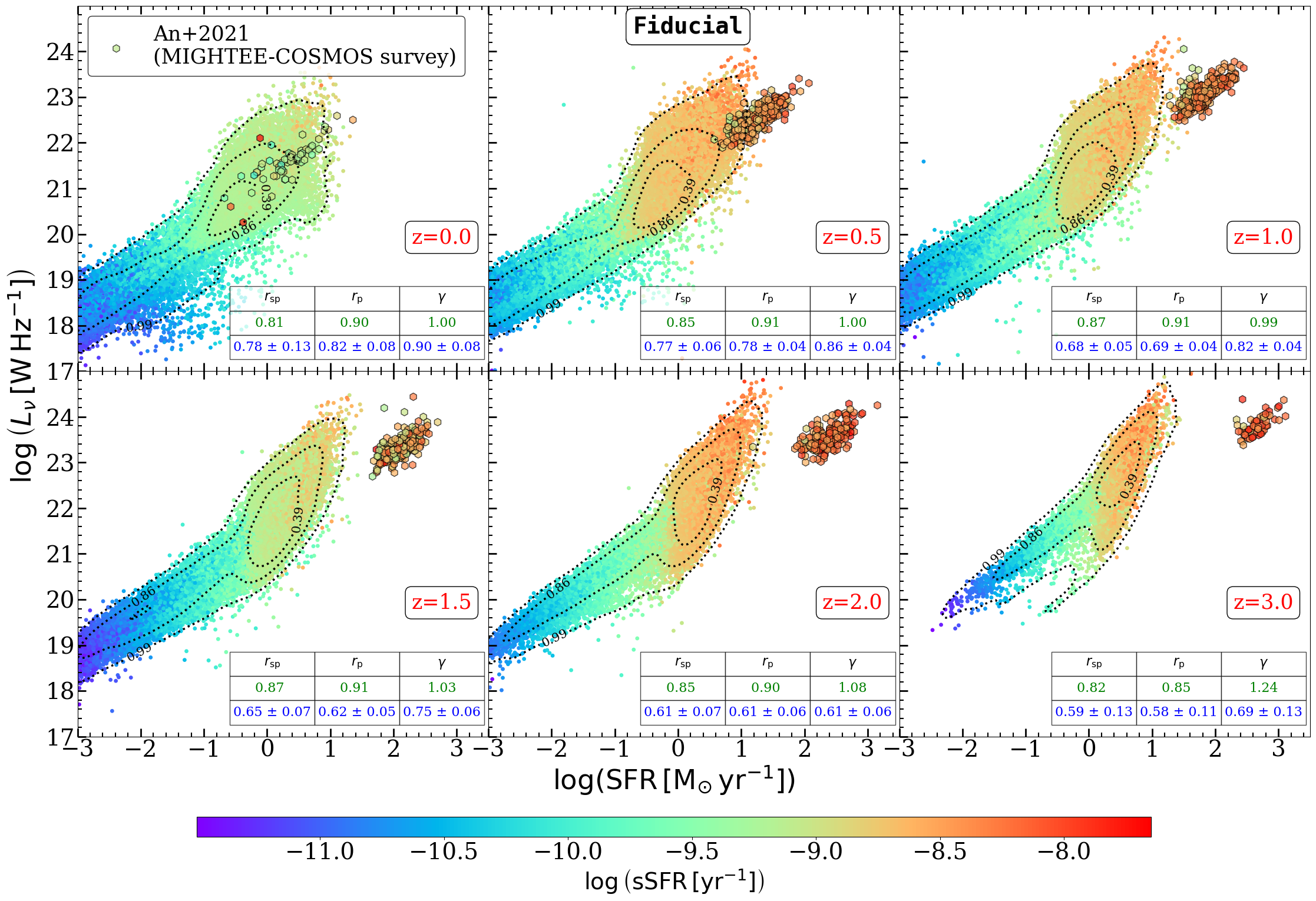}
    \caption{The correlation between $\Lspec$ and $\SFR$ as shown in Fig.~\ref{fig: SI vs SFR 4.8 b}, 
    at different redshifts up to $z=3$. 
    The observational data are taken from the MIGHTEE-COSMOS survey \citep{An+21} at the observational frequency $1.3\GHz$. 
    To compare with the observations, we first subtract the thermal contribution from the observed flux 
    and then calculate the rest-frame synchrotron luminosity 
    (see Section~\ref{sec:data} for details). 
    Each panel therefore corresponds to the rest-frame frequency $\nu_{\rm rest} = 1.3\,(1+z)\GHz$ 
    for both the simulated and observed galaxies. 
    Since the observational sample spans a wide redshift range, 
    we take the observational data for the redshift interval $\Delta z = 0.2$ in each panel,
    i.e., $z=0$--$0.2$, $0.4$--$0.6$, $0.9$--$1.1$, $1.4$--$1.6$, $1.9$--$2.1$ and $2.9$--$3.1$.}
    \label{fig: L_SFR_z_evolve}
\end{figure*}
\begin{figure*}
    \centering
    \includegraphics[width=0.80\textwidth]{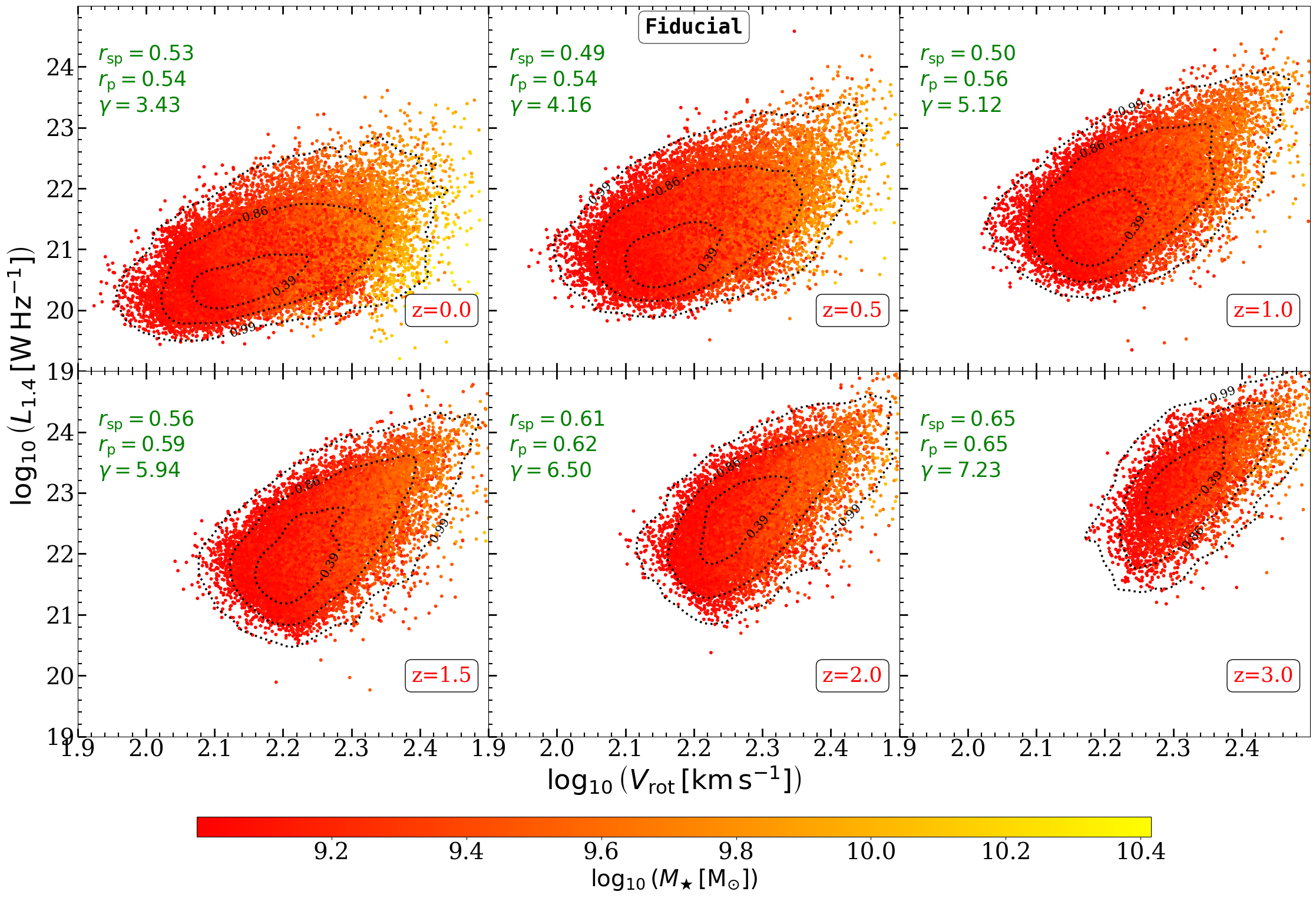}
    \caption{The redshift evolution of the correlation between $\Lspec$ and the rotational velocity $V\rot$ 
    corresponding to the flat part of the rotation curve, 
    as in Fig.~\ref{fig:Lum_I_Vrot_4.8 b}.
    }
    \label{fig:8}
\end{figure*}
\subsection{Correlation between $\Lspec$ and $V_{\rm rot}$}\label{sec:Vrot}
Motivated by the empirical correlation found 
between $\Lspec$ and rotation speed in the flat part of the rotation curve $V\rot$ in \citetalias{Tabatabaei+16},
we now investigate to what extent such a correlation arises in our theoretical model.
While the galactic differential rotation plays a primary role in the induction of the large-scale magnetic field, 
the large-scale field turns out to be subdominant in the synchrotron luminosity compared to the small-scale field. 
Moreover, if the small-scale magnetic field is artificially set to zero (leaving the large-scale magnetic field alone),
then the correlation between $\Lspec$ and $V\rot$ obtained is too weak to explain the observational data,
as discussed in Section~\ref{sec:relative}.
Therefore, if $L_\nu$ and $V\rot$ are correlated in the model, 
this cannot be attributed to the stretching of magnetic fields by differential rotation.
The small-scale magnetic field also does not depend on the galactic differential rotation in our model
(but see Section~\ref{sec:anisotropy} for a discussion). 
For these reasons, our model does not predict a \textit{causal} correlation between $L_\nu$ and $V\rot$, 
but this does not preclude a non-causal correlation.

As a measure of the rotation speed in the model galaxies, 
we use the mean rotation speed between $1.5\,r_{1/2}$ and $2.5\,r_{1/2}$ (for details see Appendix~\ref{App: Flat rotation curve}).
Figure~\ref{fig:Lum_I_Vrot_4.8 b} shows the relation between $L^\mathrm{nt}_I$ and $V\rot$ at $z=0$ and $1.4\GHz$, 
the \texttt{Fiducial} model.
Similar plots are shown in Fig.~\ref{fig:fig5} at $144\MHz$, and $4.8\GHz$.
We also include the correlation coefficients and fit parameters for $10.7\GHz$ in Table~\ref{tab:obs_data_corr}.
Our models show a strong correlation between $\Lspec$ and $V\rot$
for the actively star-forming galaxies alone ($\sSFR >10^{-10.4}\,\rm{yr^{-1}} $),
and we select only such galaxies in both the models and observational data. 
We find that more massive galaxies tend to exhibit higher rotation velocities and greater radio continuum luminosities.
The correlation between $\Lspec$ and $V\rot$ is quite strong for model \texttt{J24} (see Fig~\ref{fig:Lum_I_Vrot_4.8 a} in Appendix~\ref{App: Alt_models}), with $r\Pe = 0.84$ and $r\Sp = 0.85$, 
whereas for the observational data the correlation is moderate with $r\Pe = 0.63 \pm 0.14$ and $r\Sp = 0.60 \pm 0.18$. 
On the other hand, \texttt{Fiducial} model produces correlation coefficients of $r\Pe = 0.53$ and $r\Sp = 0.54$,
which are in better agreement with observations.
The slope is consistent with observation for both models, although for \texttt{J24} it agrees slightly better than for \texttt{Fiducial}.

Thus, the model reproduces rather well the $\Lspec$--$V\rot$ correlation seen in the observational data
despite the lack of any causal effect of the galactic differential rotation on the luminosity.
We suggest an explanation for this in Section~\ref{sec:explanation}.

\subsection{Redshift evolution}\label{sec:redshift_evolution}
Figure~\ref{fig: L_SFR_z_evolve} illustrates the redshift evolution of the relationship between $\Lspec$ and  $\SFR$, 
up to $z = 3$, for the \texttt{Fiducial} model. 
The observational data are taken from the MIGHTEE-COSMOS survey \citep{An+21} at the observational frequency $1.3\GHz$. 
After the adjustments described in Section~\ref{MCD},
each panel refers to the rest-frame frequency $\nu_{\rm rest} = 1.3\,(1+z)\GHz$ 
for both the simulated and observed galaxies. 
Since the observational sample spans a wide redshift range, 
we combine the observational data for the redshift interval $\Delta z = 0.2$ in each panel. 
We also computed the $\sSFR$ for both the simulated and observational data.

A strong correlation between $\Lspec$ and $\SFR$ 
($r\Pe, r\Sp \gtrsim 0.8$) is maintained across all redshifts, and the slope increases with the redshift.
This variation appears to be driven by changes in the relative populations of actively star-forming and quiescent galaxies.
At higher redshifts, the fraction of actively star-forming galaxies is higher.
This can be attributed to the higher galaxy merger rates which trigger intense star formation. 
This increases the turbulent speed (see equation~\ref{eq: sfr-vt}),
which makes the small-scale magnetic field stronger for $\SFR > 1\Msunyr$. 

While \textsc{galform} predicts $\SFR$ values that are in a good agreement with observations at low redshifts, 
there is a deficiency of galaxies with high SFR ($\gtrsim 10\Msunyr$) at high redshifts
\citepalias{Lacey+16}. 
However, the SFR of the observed galaxies is derived from the total IR luminosity, 
which might overestimate the SFR and also can fail to detect galaxies with low SFR \citep{Katsianis+21b,Traina+26}.
More importantly, selection effects limit the observational sample to the brightest galaxies in radio and IR,
and this bias is stronger at a higher redshift.
These issues are discussed further in Section~\ref{sec: SFR_limitation}. 

Figure~\ref{fig:8} presents the redshift evolution of the relationship between $\Lspec$ and $V\rot$ from $z = 3$ to $z=0$. 
This analysis is restricted to actively star-forming galaxies (similar to Fig.~\ref{fig:Lum_I_Vrot_4.8 b}), 
selected according to the specific star formation rate ($\sSFR$) thresholds determined at the $68\%$ confidence level 
of the $\sSFR$--$M_\star$ main-sequence population (see Appendix~\ref{app:sSFR} for details).
The strength of the correlation between $\Lspec$ and $V\rot$ is fairly constant with the redshift. 
However, the slope increases strongly with redshift, 
from about $4$ at $z\approx0$ to $9.5$ at $z\approx3$.
As $z$ increases, $\SFR$ increases, leading to more galaxies with $\SFR\gtrsim 1\Msunyr$ 
and thus high turbulent speeds according to equation~\eqref{eq: sfr-vt}, 
resulting in higher magnetic field strengths and synchrotron luminosities.
\subsection{Comparing contributions from large-scale and small-scale magnetic fields}\label{sec:relative_app}
\begin{figure*}
\begin{center}
     \includegraphics [width=0.80\textwidth]{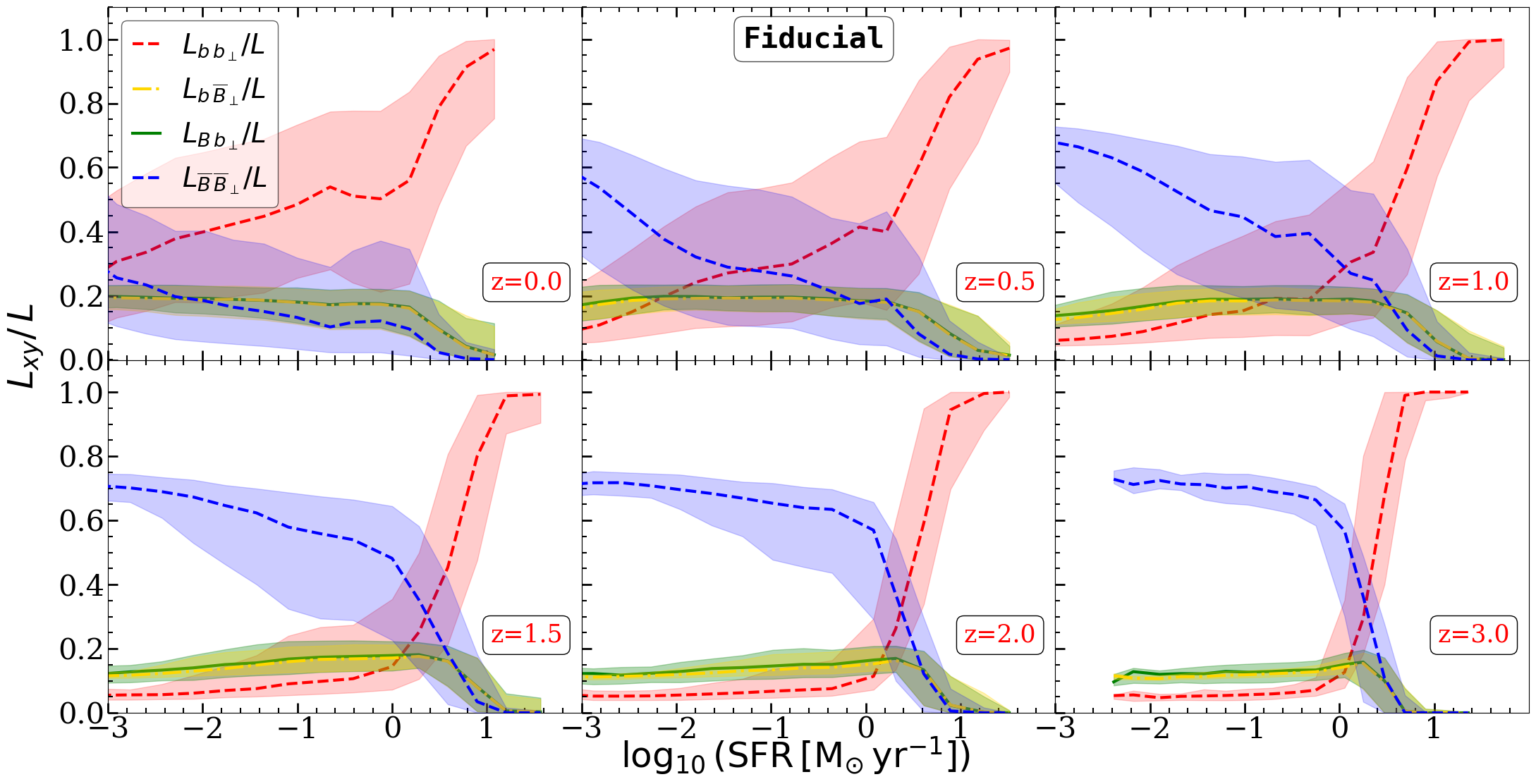}
     \caption{The redshift evolution of the ratios 
     of the specific luminosities contributed by different terms ($b^2 b_{\perp}^2$, 
     $\ol{B}^2 b_{\perp}^2$, $b^2\ol{B}_{\perp}^2$, 
     and $\ol{B}^2\ol{B}_{\perp}^2$) to the total synchrotron specific luminosity, 
     plotted as a function of the star formation rate $\SFR$. 
     The dashed red, solid green, dash-dotted yellow, 
     and dashed blue lines indicate the medians of $L_{b b_{\perp}}/L$, 
     $L_{\ol{B} b_{\perp}}/L$, $L_{b \ol{B}_{\perp}}/L$, 
     and $L_{\ol{B} \ol{B}_{\perp}}/L$, respectively. 
     The shaded region represents the 68th percentile range of the galaxy population.
     The frequency $\nu$ is dropped in the notation because the ratios are independent of frequency. 
     \label{fig:L_relative_contb}
     }
\end{center}
\end{figure*}
The total synchrotron flux density includes the volume integral of $B^2B_{\perp}^2$, 
which contains the terms $\ol{B}^2\ol{B}_{\perp}^2$, $\ol{B}^2b_{\perp}^2$, $b^2\ol{B}_{\perp}^2$ and $b^2b_{\perp}^2$. 
Figure~\ref{fig:L_relative_contb} shows the redshift evolution of the individual relative contributions 
of these terms to the specific luminosity.
In \texttt{Fiducial}, for low redshift $(z\lesssim1)$, 
the term associated with the small-scale field alone, 
$L_{bb_{\perp}}$, 
provides the dominant contribution. 
The contribution of the large-scale field, $L_{\ol{B}\ol{B}_{\perp}}$, increases with increasing redshift. For $z\gtrsim 1$, $L_{\ol{B}\ol{B}_{\perp}}$ 
dominates at low to moderate SFR.
At higher SFR ($\SFR \gtrsim 1\Msunyr$), 
the turbulent speed increases with the SFR as a power law (equation~\ref{eq: sfr-vt}), 
which enhances the small-scale magnetic field. 
However, this also increases the midplane pressure, which in turn increases the large-scale field strength 
because of the larger dynamo number, i.e., more intense large-scale dynamo action. 
But the overall strength of the small-scale field is higher compared to the large-scale field for $\SFR \gtrsim 1\Msunyr$, 
up to $z\sim 3$.
As a result, for high SFR, the small-scale field dominates over the large-scale field across all redshifts.
The terms $L_{\ol{B}b_{\perp}}$ and $L_{b\ol{B}_{\perp}}$ 
contribute almost equally throughout the redshift range, with small differences due to inclination effects. In Appendix~\ref{sec:relative}, the $\Lspec-\rm SFR$ and $\Lspec-V_{\rm rot}$ correlations for only large-scale and small-scale field contributions are shown in Figures~\ref{fig:L_SFR_LS_SS} and \ref{fig:L_Vrot_LS_SS}, respectively.
\subsection{Revisiting the radio luminosity function}
\begin{figure*}[t!]
\centering
	\includegraphics[width=0.92\textwidth]{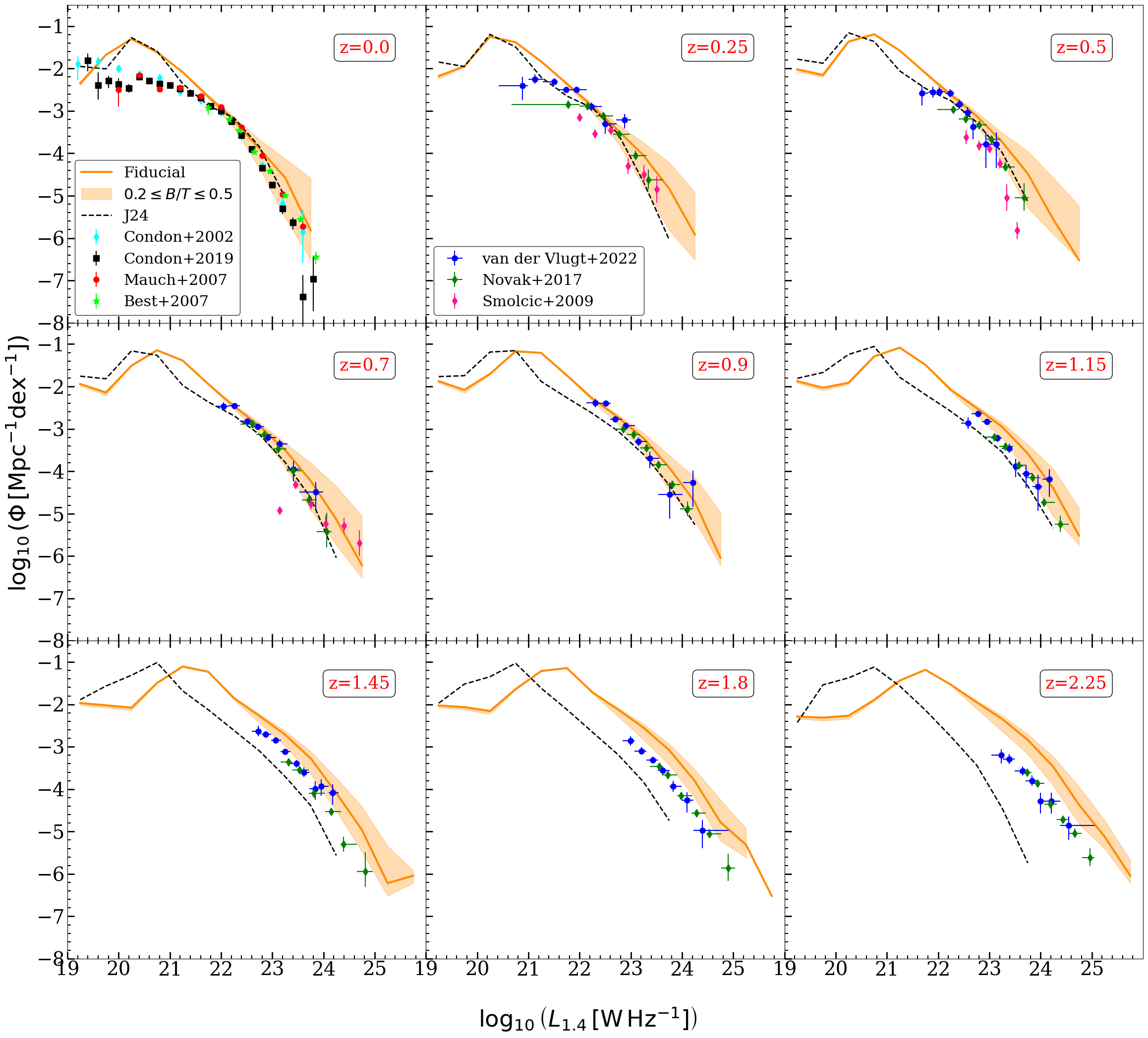}
    \caption{Redshift evolution of radio luminosity function (RLF) of SFGs at 1.4 GHz rest frame frequency. The orange curve shows the prediction from our \texttt{Fiducial} model, while the black dotted curve represents the prediction from the \texttt{J24} model. The orange shaded region indicates the prediction of our \texttt{Fiducial} model for $0.2\leq B/T\leq 0.5$. The observational data are taken from \citet{Condon+02} (cyan diamonds), \citet{Condon+19} (black squares), \citet{Mauch&Sadler07} (red circles), \citet{Best+05} (lime stars), \citet{vanderVlugt+02} (blue circles), \citet{Novak+17} (green diamonds) and \citet{Smolcic+09} (pink diamonds). }
    \label{fig: RLF_z_evolve}
\end{figure*}
The radio luminosity function (RLF) is a key tool for understanding the evolution of radio emission 
from star-forming galaxies across cosmic time. 
A detailed analysis of the RLF of SFGs using \textsc{magnetizer} was presented in \citetalias{Jose+24}. 
In Figure ~\ref{fig: RLF_z_evolve}, we predict the RLF
using our \texttt{Fiducial} model (orange solid), and compare it with the prediction of \citetalias{Jose+24} 
(Model \texttt{J24}, black dotted), 
and with various observational data. 
To exclude the contamination from radio-loud AGN, 
we restrict $B/T$ to be less than or equal to $0.4$ in \texttt{Fiducial} 
(for details see Section~\ref{sec:selection}). 
In Fig.~\ref{fig: RLF_z_evolve}, 
we show the range $0.2$--$0.5$ for this upper threshold using orange shading
(there is no such threshold for \citetalias{Jose+24}).
For redshifts up to $z\sim 1$ both \texttt{J24} and \texttt{Fiducial} show good agreement with observational data. 
However, 
\texttt{Fiducial} starts to overpredict the RLF 
above $z\gtrsim1.8$, whereas \texttt{J24} starts to underpredict it. 
Nevertheless, it is encouraging that plausible variations of parameters in \textsc{magnetizer} 
result in models that bracket the observational data.
Note that both \texttt{J24} and \texttt{Fiducial} show a local maximum near the faint end of the RLF at every redshift. 
This happens because \textsc{galform} predicts a large population of low mass galaxies. 
This is also reflected at the faint end of the IR luminosity function predicted by \citetalias{Lacey+16}.

\subsection{Correlations between $B$ and $\SFR$, $M_\star$ and $\sSFR$}
\begin{figure*}
\centering
    \includegraphics[width=0.9\textwidth]{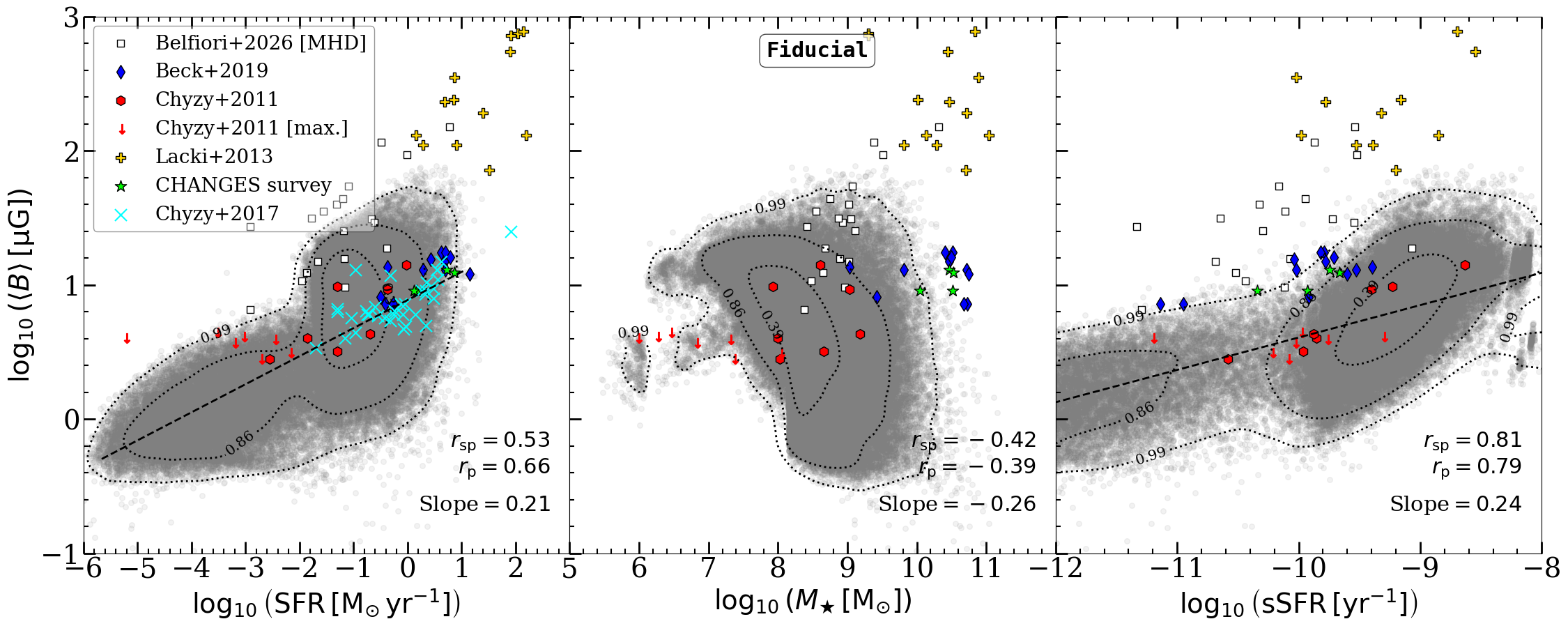}
    \caption{Correlations between the average magnetic field strength $\langle B \rangle$ 
    and $\SFR$ (left), 
    stellar mass $M_\star$ (middle) and $\sSFR$ (right). 
    The grey circles represent the fiducial model, while the blue diamonds, 
    red hexagons/downward arrows, yellow pluses, cyan crosses  and green stars correspond to observations
    from \citet{Beck+19}, \citet{Chyzy+11}, \citet{Lacki&Beck13}, \citet{Chyzy+17}
    and the CHANGES survey \citep{Mora-Partiarroyo+19,Stein+19,Stein+20,Heald+22}, respectively.
    The yellow crosses refer to localized starburst regions rather than whole galaxies. 
    The red downward arrows indicate upper limits. The data from \citet{Chyzy+17} are not included in the middle and right panels because stellar mass estimates are unavailable.
    The black dashed lines show the linear fits to our fiducial model, and the black dotted contours 
    represent the $39\%$, $86\%$ and $99\%$ confidence levels of the 2D kernel density estimate. 
    }
    \label{fig:B_SFR_Mstar_sSFR}
\end{figure*}
Alternatively, scaling relations between some measure of the average magnetic field strength
and various global observables can be studied, 
though inferring the magnetic field strength from observations requires considerable modelling \citep{Beck+19}.
\citet{Belfiori+26} compiled data of magnetic field strength and other galactic parameters from the literature 
to explore the relationships between magnetic field strength and $\SFR$, $M_\star$ and $\sSFR$.
The empirical relations are shown in Fig.~\ref{fig:B_SFR_Mstar_sSFR}, 
with observational data for nearby galaxies represented by the colored symbols.
The data shows the average magnetic field strength derived from the observed radio intensity assuming equipartition 
between the magnetic field and cosmic ray energy densities \citep{Beck&Krause05,Lacki&Beck13}, 
which gives $B\propto (\Lspec/V)^{2/(s+5)}$, 
where $V$ is the volume of the galaxy and $s$ is the spectral index of cosmic ray electrons.
Thus, to compare our models with the observational data, we use
\begin{equation}\label{eq:B_avg}
    \langle B\rangle = \left(V^{-1}\int_V B^4(\bm{r})\, \di^3\bm{r}\right)^{1/4},
\end{equation}
where integration extends up to the maximum disc radius $r_{\rm d}$ and the gas scale height $h_{\rm d}$. 
The yellow crosses represent \textit{localized} starburst regions of very high $\SFR$ and magnetic field strength 
from \citet{Lacki&Beck13}, and should not be directly compared with our model predictions.
The predictions of Model \texttt{Fiducial} are shown as grey circles with probability density contours
in Fig.~\ref{fig:B_SFR_Mstar_sSFR}. 
The observational data also include galaxies with stellar masses below $10^8\,\rm M_{\odot}$.
To obtain an appropriate sample for comparison with these data, 
we selected $10^5$ model galaxies from $z=0$ to $z=2$ from \texttt{Fiducial}
without imposing a minimum stellar mass or SFR cutoff.

These relations were explored by \cite{Belfiori+26} using the median magnetic field strength 
across all resolution elements from adaptive mesh refinement (AMR) MHD simulations. 
From our model, we obtain a strong correlation between $\langle B\rangle$ and $\SFR$, with a slope of $0.21$, 
in fairly good agreement with the observations 
and in agreement with the slope of $0.28\pm0.08$ found for the AMR MHD simulations.
For the $\langle B\rangle$--$\sSFR$ relation, we obtain the slope $0.24$, 
which is reasonably consistent with the observational trend 
and marginally consistent with the simulation result of $0.36\pm0.10$. 
Consistent with the observations, 
we do not find a significant correlation between the magnetic field strength and stellar mass,
whereas \citet{Belfiori+26} obtain a correlation between these quantities from their simulations.
Galaxies in their sample are selected at $z=3$ whereas we randomly select $10^5$ galaxies from $z=0$--$2$.

In any case, 
it is encouraging that strikingly different theoretical models (ours and that of \citealt{Belfiori+26})
produce the $\langle B \rangle$--$\SFR$ and $\langle B\rangle$--$\sSFR$ relations similar to one another 
and fairly consistent with trends seen in the observational data.
The model of \citet{Belfiori+26} is intrinsically three-dimensional, 
includes detailed multi-physics implementations, and allows for various types of feedback neglected in our model.
On the other hand, the dynamo found in their simulations
may not be adequately resolved given that the minimum cell size ($20\pc$) 
is comparable to the correlation scale estimated for interstellar turbulence \citep[e.g.][]{Chamandy+Shukurov20}. 
In our model and that of \citet{Belfiori+26}, the small-scale magnetic field dominates the magnetic field strength. 
We model the small-scale magnetic field strength as a fixed fraction $f_b = 0.8$ 
of the local turbulence equipartition strength $(4\uppi\rho)^{1/2}v\turb$.
Hence, it would be instructive to study to what extent this assumption is consistent with their and other MHD simulations.

\section{Discussion}\label{sec:discussion}
The galactic dynamo population model \textsc{magnetizer} 
coupled with the synchrotron emission model reproduces the observational correlations  
between $\Lspec$ and $\SFR$, and between $\Lspec$ and $V\rot$ found in nearby galaxies.
What explains these correlations?

\begin{figure}
\centering
	\includegraphics[width=\columnwidth]{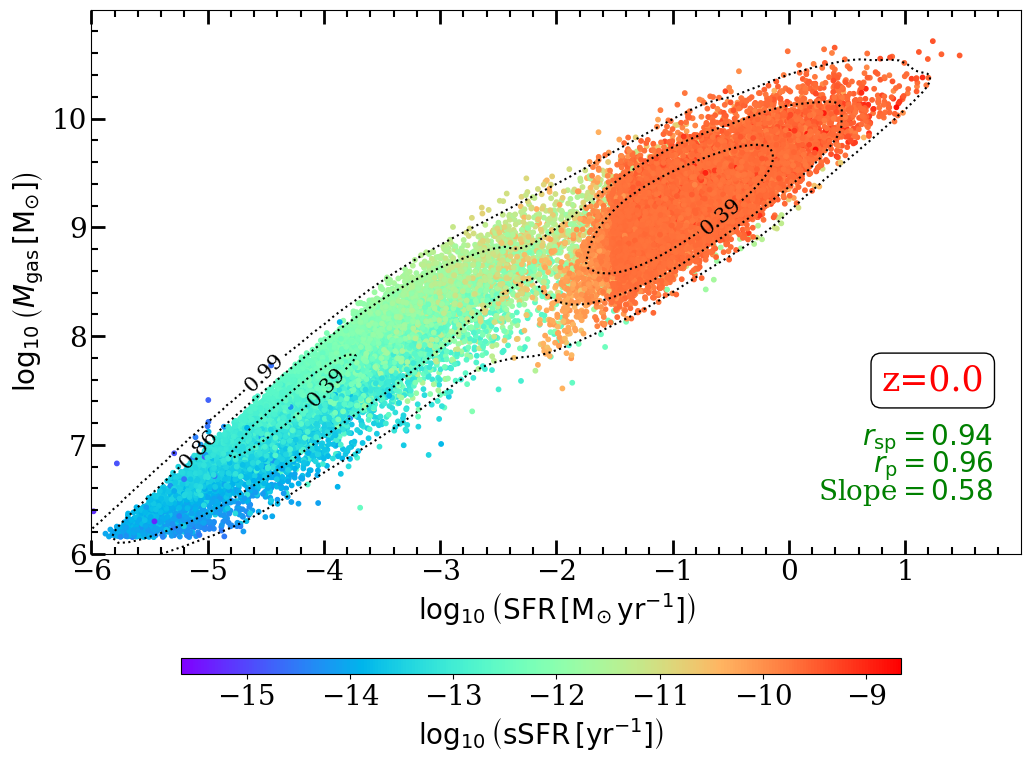}
    \caption{The correlation between the star formation rate $\SFR$ and the total disc gas mass at the redshift $z = 0$. 
    This correlation helps to explain the correlation seen between $\Lspec$ and $\SFR$.}
    \label{fig:Mgas_vs_SFR}
\end{figure}
\begin{figure}
    \centering
	\includegraphics[width=\columnwidth]{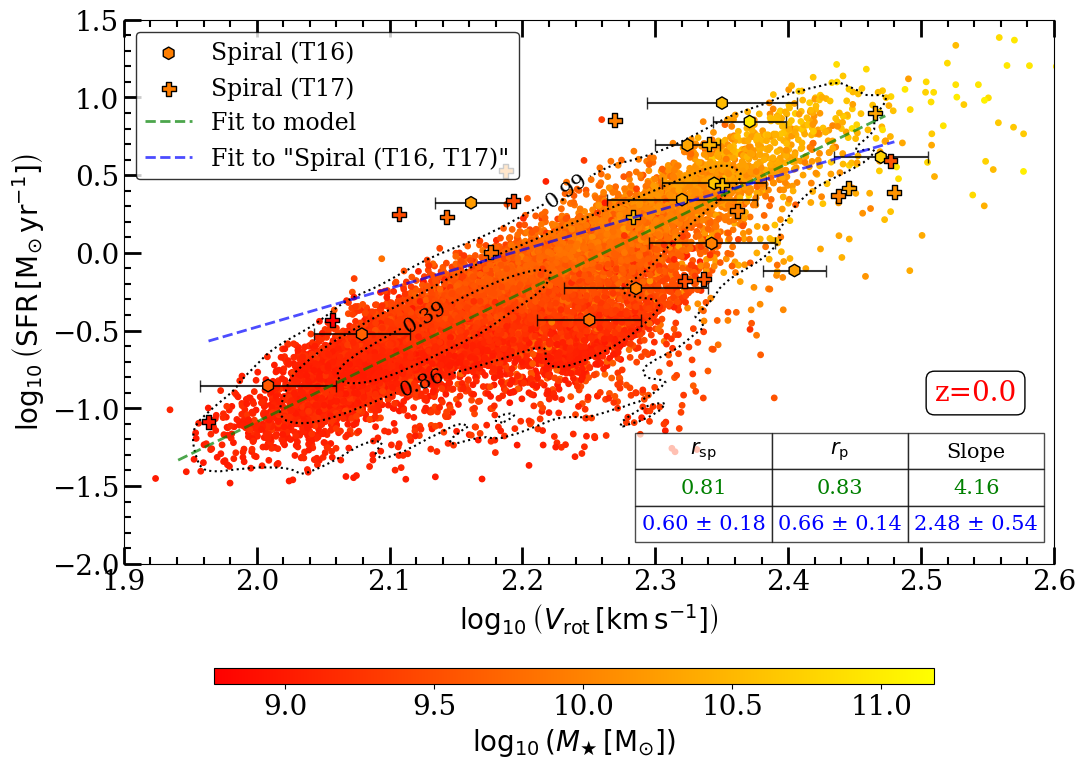}
    \caption{The correlation between the star formation rate $\SFR$ and the rotational 
    speed $V\rot$ 
    corresponding to the flat part of the rotation curve for actively star-forming galaxies 
    with specific star formation rates $10^{-10.4}\yr^{-1}$ at the redshift $z = 0$. 
    The color bar represents the stellar mass $M_\star$ of the galaxies. 
    Simulated galaxies are shown as circles, 
    while the observational spiral galaxy samples from \citetalias{Tabatabaei+16} and \citetalias{Tabatabaei+17} 
    are shown as pluses and hexagons with error bars.}
    \label{fig:SFR_vs_Vrot}
\end{figure}
\begin{figure}
    \centering
	\includegraphics[width=\columnwidth]{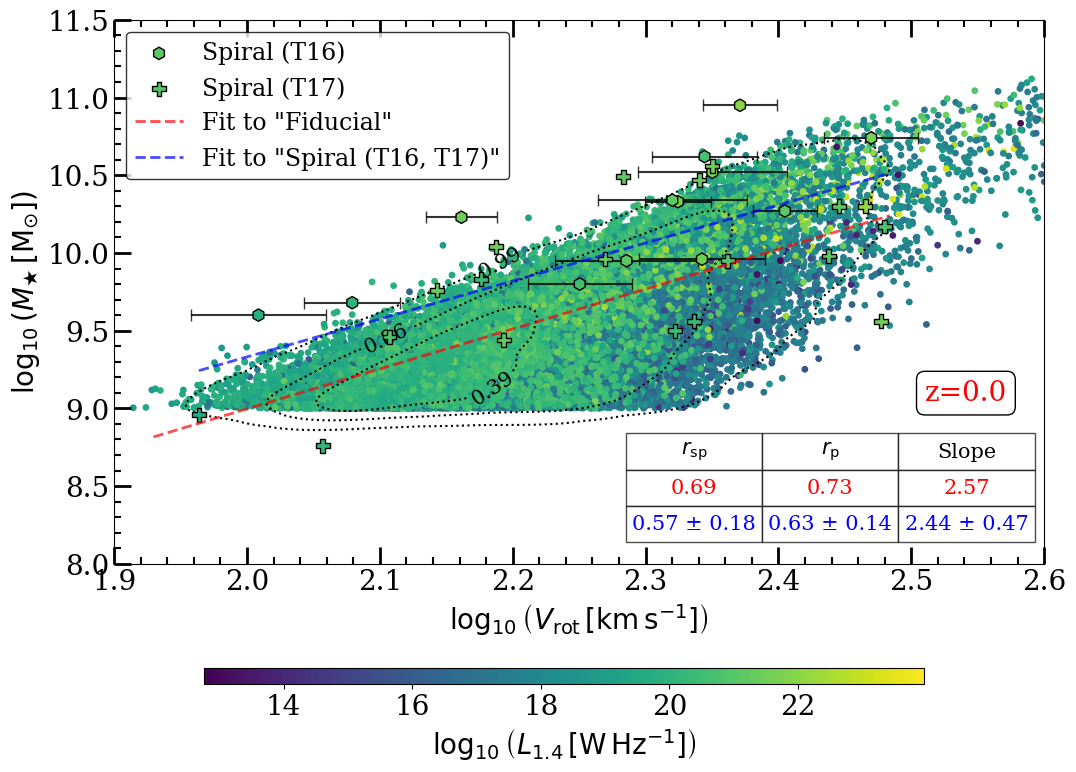}
    \caption{
    The correlation between the stellar mass $M_\star$ and the rotational velocity $V\rot$ 
    corresponding to the flat part of the rotation curve for actively star-forming galaxies 
    with specific star formation rates $\sSFR > 10^{-10.4}\yr^{-1}$ at the redshift $z = 0$. 
    The color bar represents the total synchrotron specific luminosity $\Lspec$ at 1.4 GHz. 
    Simulated galaxies are shown as circles, 
    while the observational spiral galaxy samples from \citetalias{Tabatabaei+16} 
    and \citetalias{Tabatabaei+17} 
    are shown as pluses and hexagons with error bars. The lower stellar mass threshold for the simulated galaxies is $10^9\,\rm M_{\odot}$.}
    \label{fig:Mstar_vs_Vrot}
\end{figure}

\subsection{Physical basis for the correlations}\label{sec:explanation}
Under the assumption of the local equipartition at scales 
of a few hundred parsecs, 
the synchrotron emissivity scales approximately as $B^4$,
where $B$ is the local magnetic field strength.
Both small-scale and large-scale magnetic fields saturate at values proportional to the local equipartition field strength, 
$B\eq = (4\uppi \rho)^{1/2} v\turb$,
so $B$ primarily depends on the local gas density $\rho(r)$ and the turbulent speed $v\turb$,
which is here assumed to be independent of position within a given galaxy and is given by equation~\eqref{eq: sfr-vt}. 

The correlation between $\Lspec$ and $\SFR$ holds over a wide range of $\SFR$s. 
However, $v\turb=15\kms$ is constant for $\SFR\le1\Msunyr$ in our model.
Therefore, the variation of $v\turb$ with $\SFR$, which only occurs for $\SFR>1\Msunyr$, 
cannot explain the observational correlation.
This leaves the gas density as the remaining driver.
However, the specific luminosity is an integrated quantity over the entire volume of the galaxy,
so it is more meaningful to consider the relationship with the total 
(sum of the diffuse and molecular) disc gas mass $M\gas$ rather than the local gas density.
Indeed, $M\gas$ and $\SFR$ are strongly correlated, as shown in Fig.~\ref{fig:Mgas_vs_SFR}.
Such a correlation between molecular gas mass and $\SFR$ 
has been found observationally \citep[e.g.,][]{Feldmann+20,Colombo+25}. 
In Fig.~\ref{fig:Mgas_vs_SFR}, the correlation we obtain is steeper for quiescent galaxies 
than for actively star-forming galaxies, 
which is also observed by \citet{Colombo+25} for molecular gas 
(we note that in \textsc{galform}, stars are assumed to form out of the molecular gas). 
Therefore, the primary driver of the $\Lspec$--$\SFR$ correlation in our model 
is the underlying star formation prescription of the galaxy formation model, rather than the turbulence prescription.
However, for $\SFR > 1\Msunyr$, $v\turb$ depends on $\SFR$ as a power law (equation~\ref{eq: sfr-vt}). 
This makes the correlation much steeper for galaxies with $\SFR > 1\Msunyr$, 
which is much more evident at high redshifts. 

We also found, in Section~\ref{sec:Vrot}, 
that there is a strong correlation between $\Lspec$ and $V\rot$
for \textit{actively} star-forming galaxies.
This correlation appears to originate from the strong correlation between $V\rot$ and $\SFR$, 
which we show in Fig.~\ref{fig:SFR_vs_Vrot} for \texttt{Fiducial}.
The plot shows good agreement with the observational data, 
although the model underestimates the amount of scatter.
The correlation between $V\rot$ and $\SFR$ results from the dependence of $V\rot$ on $M_\star$,
which is the well-known stellar mass Tully--Fisher relation \citep[e.g.,][]{McGaugh+00,McGaugh&Schombert2015}. 

In Fig.~\ref{fig:Mstar_vs_Vrot}, 
we plot the correlation between $M_\star$ and $V\rot$ from our model at $z=0$ 
for actively star forming galaxies only, with lower mass threshold $10^9\Msun$.
However, this strong correlation holds for the entire disc galaxy population. 
The color bar represents the synchrotron specific luminosity of the galaxies at $1.4\,\rm GHz$. 
We obtain a strong correlation ($r_{\rm p},\, r_{\rm sp}\gtrsim 0.7$) between these quantities 
in a fairly good agreement with observational data from \citetalias{Tabatabaei+16} and \citetalias{Tabatabaei+17}. 
However, the slope of this relation in our model for the whole population of galaxies (about $3.6$) 
is lower than the empirical estimates reported by \citet{McGaugh&Schombert2015} (about $4.5$--$5$). 
Combining this with the correlation between $M_\star$ and $\SFR$
(the star-forming main sequence; e.g.,~\citealt{Popesso+23}; \citealt{Kalinova+21}) 
for actively star-forming galaxies, we obtain the correlation between $V\rot$ and $\SFR$. 
The star-forming main sequence is evident in both the data and model by noting the color gradient
of the points with respect to the $\SFR$ in Fig.~\ref{fig:SFR_vs_Vrot}.
Both the $\SFR$--$V\rot$ and $M_\star$--$V\rot$ correlations 
remain throughout the redshift range $0\le z\le3$, 
and the $\Lspec$--$\SFR$ and $\SFR$--$V\rot$ correlations are together responsible for the correlation between $\Lspec$ and $V\rot$. 

\subsection{Robustness of the model}
\subsubsection{Frequency dependence}
In our model, 
the frequency in the source rest frame $\nu$ only affects the overall magnitude of the emissivity (see equation~\ref{eq:5}).
As a result, the power law exponents in the scaling relations are independent of the frequency. 
This model prediction is generally borne out in the data, 
where the values of $\gamma$ for both the $\Lspec$--$\SFR$ and $\Lspec$--$V\rot$ power-law fits
are consistent across different $\nu$ values within the estimated uncertainties (Table~\ref{tab:data}).
\subsubsection{Sensitivity to the dynamo parameters}\label{sec:sensitivity_to_parameters}
The model has two adjustable dynamo parameters, $f_b$ and $R_\kappa$.
The strength of the large-scale magnetic field increases with $R_\kappa$ 
while the strengths of both the large- and small-scale fields increase with $f_b$.
Increasing $R_\kappa$ enhances the diffusion of magnetic helicity density, 
which supports dynamo action by reducing the Lorentz force acting on the turbulent interstellar gas. 
Increasing $f_b$ directly enhances the small-scale magnetic field (equation~\ref{eq:1}), 
which raises the midplane pressure (Appendix~\ref{app: midplane pressure}) and consequently increases the gas scale height,
further strengthening the large-scale magnetic field.\footnote{The magnetic contribution 
to the pressure is proportional to $b\rms^2\propto f_b \rho v\turb^2$.
To get a rough idea for why this increases $h\disc$, 
suppose that gravity is dominated by a thin stellar disc. 
Then vertical hydrostatic balance gives $\del P/\del Z \simeq \rho v\turb^2/h\disc\simeq -\rho g$, 
with $g\approx\const$, and hence $h\disc\propto v\turb^2$ and $h\disc$ increases with $f_b$.
For details see appendix~A3 of \citetalias{Rodrigues+19}.
Further, the saturation strength of the large-scale field increases with the 
dynamo number $D$, which is proportional to $h\disc^2/v\turb^2$ \citep{Chamandy+14b}.}
We also find that for $f_b\ll1$, the small-scale field strength decreases but 
dominates the luminosity relative to the large-scale field even more 
than it does for $f_b\simeq 1$. 
This is because reducing $f_b$ lowers the midplane pressure by reducing its magnetic contribution, 
which in turn reduces the gas scale height.
This leads to a stronger reduction in the large-scale magnetic field strength than in the small-scale magnetic field.
To the extent that the small-scale magnetic field dominates the total synchrotron emission (see Section~\ref{sec:relative_app}), as is often inferred for nearby galaxies \citep{Beck+19},
the level of agreement between the models and observations is not sensitive to $R_\kappa$.
We have checked that a larger $f_b$ leads to a higher overall magnitude of $\Lspec$,
but the degree of correlation and slope $\gamma$ are not sensitive to $f_b$.


Table~\ref{tab:dependence_fb} illustrates the dependence of the $\Lspec$--$\SFR$ relation on $f_b$, 
presenting the correlation coefficients ($r_{\rm sp}$ and $r_{\rm p}$), 
the slope $\gamma$ and the normalization factor at $1.4\GHz$.
It is evident that values of $f_b$ in the range $0.8$–$1.0$ 
provide a reasonable agreement with the observational data across different frequencies. 
We adopt $f_b = 0.8$ for \texttt{Fiducial} as this value leads to a good overall agreement with the observational data. 
\begin{table}
    \centering
    \footnotesize
    \caption{Spearman's rank coefficient, Pearson correlation coefficient, the power law index (slope) $\gamma$, and the normalized luminosity at $1.4\,$GHz for the $\Lspec$--$\SFR$ correlation for different turbulence equipartition fractions $f_b$. The row with $f_b = 0.8$ corresponds to the fiducial model, and other rows to runs that differ from \texttt{Fiducial} only in the value of $f_b$.}
    \label{tab:dependence_fb}
    \begin{tabular*}{\columnwidth}{@{\extracolsep{\fill}} c c c c c }
        \hline\hline
        $f_b$ & $r\Sp$ & $r\Pe$ & $\gamma$ & $\log\left(\dfrac{L_{0,\,1.4}}{\rm W\,Hz^{-1}}\right)$ \\
        \hline
        $1$   & 0.82 & 0.91 & 1.02 & 20.39 \\
        $0.9$ & 0.81 & 0.91 & 1.01 & 20.25 \\
        $0.8$ & 0.81 & 0.90 & 1.00 & 20.09 \\
        $0.5$ & 0.76 & 0.85 & 0.93 & 19.52 \\
        \hline
    \end{tabular*}
\end{table}
\begin{table}
    \centering
    \footnotesize
    \caption{Spearman's rank coefficient, Pearson correlation coefficient, the power law index (slope) $\gamma$ and the normalized luminosity at 1.4 GHz of the $\Lspec$--$\SFR$ correlation for different bulge-to-total $(B/T)$ cutoffs. Model~\texttt{Fiducial} has $B/T\le0.4$ and other rows of the table differ from \texttt{Fiducial} only in the value of this parameter.}
    \label{tab:dependence_B_T}
    \begin{tabular*}{\columnwidth}{@{\extracolsep{\fill}} c c c c c }
        \hline\hline
        $\mathrm{max}(B/T)$ & $r\Sp$ & $r\Pe$ & $\gamma$ & $\log\left(\dfrac{L_{0,\,1.4}}{\rm W\,Hz^{-1}}\right)$ \\
        \hline
        $1$   & 0.81 & 0.82 & 1.04 & 20.28 \\
        $0.8$ & 0.82 & 0.85 & 1.03 & 20.19 \\
        $0.5$ & 0.82 & 0.90 & 1.01 & 20.10 \\
        $0.4$ & 0.81 & 0.90 & 1.00 & 20.09 \\
        $0.2$ & 0.79 & 0.91 & 1.00 & 20.09 \\
        \hline
    \end{tabular*}
\end{table}
\begin{table}
    \centering
    \footnotesize
    \caption{Spearman's rank coefficient, Pearson correlation coefficient, power law index (slope) $\gamma$ and normalized luminosity at $1.4\,$GHz for $\Lspec$-$\SFR$ correlation for different stellar mass lower threshold. Model~\texttt{Fiducial} has $M_\star\geq10^{9}\Msun$ and other rows of the table show runs that are the same as \texttt{Fiducial} except for the value of this parameter.}
    \label{tab:Mstar_dependance}
    \begin{tabular*}{\columnwidth}{@{\extracolsep{\fill}} c c c c c }
        \hline\hline
        $M_{\star, \mathrm{min}}\;[\mathrm{M_{\odot}}]$ & $r_{\mathrm{sp}}$ & $r_{\mathrm{p}}$ & $\gamma$ & $\log\left(\dfrac{L_{0,\,1.4}}{\mathrm{W\,Hz^{-1}}}\right)$ \\
        \hline
        $10^8$    & 0.74 & 0.82 & 0.85 & 20.32 \\
        $10^{8.5}$ & 0.77 & 0.87 & 0.94 & 20.23 \\
        $10^{9}$   & 0.81 & 0.90 & 1.00 & 20.09 \\
        $10^{9.5}$ & 0.85 & 0.91 & 1.03 & 19.93 \\
        \hline
    \end{tabular*}
\end{table}

\subsubsection{The bulge-to-total mass ratio}
We employ an upper limit to the ratio of stellar mass in the bulge to the total stellar mass $B/T$
in order to include disc galaxies and exclude elliptical and lenticular galaxies.
Moreover, our model does not consider the radio emission from AGNs, 
so it is important to exclude galaxies whose radio emission would be dominated by AGN activity.
Galaxies with high bulge masses are more likely to host AGNs because of the presence of more massive central black holes. 
In this work, we use $B/T\le 0.4$, which is motivated by observational data (see Appendix~\ref{app:B_T_ratio}). 
However, we have explored the sensitivity to the choice of $\mathrm{max}(B/T)$ 
and the results are presented in Table~\ref{tab:dependence_B_T}.
Increasing this threshold leads to the inclusion of galaxies with higher luminosities and higher SFRs. 
Galaxies with high SFR ($\SFR\gtrsim 1\Msunyr$) tend to have larger turbulent speeds (equation~\ref{eq: sfr-vt}), 
which produces stronger magnetic fields. 
These galaxies lead to more scatter in the $\Lspec$--$\SFR$ relation at high SFR (resulting in a lower $r\Pe$),
as well as a slightly higher slope.
Overall, our results are not very sensitive to the choice of the $B/T$ threshold,
but higher thresholds lead to an overprediction of the abundance of galaxies at
the bright end of the RLF.

\subsubsection{The lower stellar mass threshold}\label{sec:Mstar_dependence}
To compare our models with the observational data, we have implemented a lower 
stellar mass threshold of $10^9\Msun$, as most spiral galaxies have stellar masses $\gtrsim 10^9\Msun$. 
However, we have explored the dependence of the $\Lspec$--$\SFR$ correlation 
on the minimum stellar mass threshold, $\min(M_\star)$, in Table~\ref{tab:Mstar_dependance}. 
We find that the correlation coefficients are not strongly affected, 
but that the slope increases significantly with increasing $\min(M_\star)$.
However, choosing values anywhere in the range $10^8$--$10^{9.5}\Msun$ 
does not lead to significant changes to our conclusions. 


\subsubsection{The model for the turbulent speed}\label{sec:vt_model_discussion}
In our model the turbulent speed is given by the piecewise expression~\eqref{eq: sfr-vt}, 
which is based on a fit to observational data for the spectral line widths \citep{Krumholz+18}. 
We performed some experiments to see how sensitive our fiducial model 
is to choices about the $v\turb$--$\SFR$ relation.
A model with a constant turbulent speed, $v\turb=15\kms$,
performs worse in explaining highly luminous galaxies with high $\SFR$. 
We also implemented a few models in which $v\turb$ is obtained 
by fitting a smooth curve to the observational velocity-dispersion--SFR correlation data compiled by \citet{Krumholz+18}. 
Such fits lead to an overprediction of the slope of the  $\Lspec$--$\SFR$ trend.

\subsection{Star formation rate}\label{sec: SFR_limitation}
Figure~\ref{fig: L_SFR_z_evolve} shows that \textsc{galform} fails to reproduce 
the population of galaxies with $\SFR \gtrsim 2\times 10^{2}\Msunyr$ at high redshift inferred from observations. 
This might be due to a combination of the following reasons:
(i)~observational selection effects may cause only the brightest, 
most vigorously star-forming galaxies to be detected even though such galaxies are rare;
(ii)~the star formation recipe of \textsc{galform} may underestimate the SFR in the high SFR regime; 
(iii)~observational inferences may tend to overestimate the SFR in the high SFR regime.

\subsubsection{Star formation prescription in \textsc{galform}}
In the \citetalias{Lacey+16} version of \textsc{galform} used in this work, 
star formation is modelled using the empirical relation of \citet{Blitz&Roso2004, Blitz&Roso2006}, 
which is based on nearby galaxies and may not be valid at a high redshifts. 
Incorporating observational constraints from high-redshift galaxies 
is thus necessary for more accurate modelling of the star formation rate.
According to \citet{Traina+26}, 
this issue of predicting too low SFR at a high redshift is common among modern galaxy formation models, 
including both hydrodynamic (HD) or magnetohydrodynamic (MHD) simulations as well as semi-analytical models (SAMs). 
They suggest two main reasons: (i)~the relatively small box sizes of HD/MHD simulations 
limit their ability to capture rare, extremely high-SFR galaxies; 
(ii)~limited numerical resolution in HD/MHD simulations prevents them 
from modelling the detailed physical processes that drive intense star formation.
SAMs, by contrast, can use much larger volumes. 
However, even SAMs remain unable to reproduce the most extremely star-forming galaxies at high redshifts ($z \gtrsim 2$), 
likely due to limitations in their treatment of the physical processes governing star formation. 

\subsubsection{SFR from high-redshift observations}
The SFR of the MIGHTEE-COSMOS galaxies is computed using the total IR ($8$--$1000\,\rm\upmu m$) luminosity \citep{Jin+18}.
This method can overestimate the SFR due to 
(i)~dust heating by older populations of stars, 
(ii)~higher emission from polycyclic aromatic hydrocarbons of distant galaxies,
(iii)~at higher redshifts, IR observations become less sensitive for detecting galaxies in low star-forming regimes, 
causing a selection bias in favour of higher star-forming regimes \citep{Madau+Dickinson16, Katsianis+21b, Traina+26}.

\subsection{Caveats}
\subsubsection{Observational selection effects}\label{sec:obs_selec_effects}
Selection effects present in the data but not the model may hamper the comparison between the model and data,
even for low-redshift galaxies.
For example, the galaxies of \citetalias{Tabatabaei+16} are isolated (not in a cluster and non-interacting) whereas 
there is no such selection in our model (indeed, there is no straightforward way to select only such galaxies in the model).
According to \citet{Hosseinirad+23}, 
the $\langle B \rangle\text{--}V_{\rm{rot}}$ correlation strengthens as the degree of isolation increases 
in the \textsc{TNG50} simulation \citep{Pillepich+19}, so this may be important.
Moreover, the observational data in \citetalias{Tabatabaei+16}, 
\citetalias{Tabatabaei+17} and \citetalias{Heesen+22} are biased in favour of high-SFR galaxies, 
which can be seen in the histogram distribution of galaxies along $\SFR$. 
We note that the distribution of galaxies in $\SFR$ found by the Sloan Digital Sky Survey (SDSS) 
\citep{Renzini&Peng15} broadly agrees with \textsc{galform} results.

\subsubsection{Cosmic ray model}
We assume local energy equipartition between cosmic rays and the magnetic field 
on scales of the \textsc{magnetizer} grid resolution (typically a few hundred parsec) in order to estimate the cosmic ray number density. 
While such an assumption is widely used, it lacks strong observational and theoretical support.
We also tested a model with a spatially constant cosmic ray number density. 
Even when allowing the normalization constant to vary freely, 
we were unable to obtain a better fit to the data than for our \texttt{Fiducial} model. 
Recent CR-MHD simulations of isolated galaxies
\citep{Chiu+25} and of a local ISM region \citep{Linzer+25} suggest that the equipartition assumption 
may be acceptable on $\!\kpc$ scales, 
but fails at scales of $10\pc$ and less \citep[see also][]{Seta+Beck19}.  
We note that the simulations of \citet{Linzer+25} do not include the effects of cosmic rays on the system.
On the other hand, test-particle simulations of cosmic rays (both protons and electrons) 
do not provide any indications of the equipartition \citep{Tharakkal+23a}. 
Simulations of the non-linear states of the Parker instability, 
where cosmic rays and magnetic fields evolve self-consistently, show that the energy densities of cosmic rays 
and magnetic fields are slightly anti-correlated on the instability scale of the order of a kiloparsec \citep{Tharakkal+23b}, 
in agreement with the analysis of the synchrotron fluctuations in the Milky Way and the galaxy M33 \citep{Stepanov+14}.
Local CR-MHD simulations with a full account of the dynamical effects of cosmic rays and magnetic fields on the ISM 
do not provide any indications of the energy equipartition between cosmic rays and magnetic fields 
(Qazi et al., 2026, in preparation).
A more realistic treatment of cosmic rays could involve deriving the cosmic ray distribution 
as an explicit solution of the transport equation of cosmic rays, e.g., as that presented by \citet{SJ26}. 

Cosmic microwave background (CMB) and starlight photons 
interact with the cosmic ray electrons of galaxies via the inverse Compton scattering. 
This is an efficient energy loss mechanism of cosmic ray electrons 
and can change both their distribution and energy spectrum.
This effect increases with increasing redshift in proportion to $(1+z)^4$
and can dominate over synchrotron losses for $z\gtrsim 2$--$3$ \citep{Murphy+09}.
However, it is difficult to incorporate this effect within the equipartition model of cosmic ray electron distribution. 
One possible way is to allow the cosmic ray spectral index $s$
to vary with redshift in a way that is consistent with observational data.
\cite{Tabatabaei+25} found a strong correlation between the nonthermal spectral index $\alpha_{\rm nt}$ 
and $z$ and between $\alpha_\mathrm{nt}$ and $\sSFR$. 
We have experimented with using these two correlations to compute $s$ in our model rather than using a constant value of $3$. 
However, we find that using the $\alpha_{\rm nt}$--$z$ correlation suggested predicts the $\Lspec$ higher 
than in the observational data at low redshifts ($z<1$), 
whereas using the $\alpha_{\rm nt}$--$\sSFR$ correlation predicts the slope of $\Lspec$--$\SFR$ 
lower than the observational one. 
The reason for this might be because, firstly, 
the galaxies used in that study are chosen from the redshift range $1.5$--$3.5$, 
so the same $\alpha_{\rm nt}$--$z$ relation may not be applicable at lower redshifts, and secondly, 
these authors use galaxies with high SFR, and the resulting $\alpha_{\rm nt}$--$\sSFR$ relation may not be appropriate
for galaxies with a lower SFR.

\subsubsection{Magnetic field pressure}
In our model, the gas scale height and density are obtained by solving the equation of vertical hydrostatic equilibrium.
The midplane pressure used in this calculation includes turbulent, thermal, small-scale magnetic and cosmic ray contributions,
but does not include the large-scale magnetic field contribution (Appendix~\ref{app: midplane pressure}).
In the current implementation of \textsc{magnetizer}, including this pressure contribution
increases the scale height $h\disc(r)$, which in turn amplifies $\overline{B}$, leading to a runaway effect. 
To avoid this instability, we set the large-scale magnetic pressure to zero.
This aspect of the model
needs to be addressed in future versions of the code.
However, given that the large-scale magnetic field pressure is subdominant,
we do not consider this to be a major problem for the current model.

\subsubsection{Galactic outflows and gaseous halo}
Another limitation is that we do not include synchrotron emission from the galactic halo, 
nor do we include outflows (winds and fountain flow) in the dynamo model.
Such outflows could affect significantly the magnetic field structure in the galactic disc
\citep[e.g.,][]{Chamandy+Taylor15}, 
and advect magnetic fields into the halo/circumgalactic medium.
Observations have shown that extended radio continuum halos are common among nearby edge-on spiral galaxies, indicating that a significant fraction of the synchrotron emission can originate in the galactic halos \citep[e.g.,][]{Irwin+12,Irwin+24, Wiegert+15}. 
Even if the magnetic field in the halo is relatively weak, 
it could contribute significantly to the synchrotron emission due to its large volume,
provided the cosmic ray electron density is sufficiently high \citep{Krause+18,Krause+20}.
These effects and others could be implemented using a 2D model (in $r$ and $Z$),
but this would require considerably more computational resources compared to the current 1D model 
(in $r$, with the no-$Z$ approximation).
Future work is needed to improve these aspects of the model.
Examples of dynamo models that incorporate the galactic halo
are \citet{Brandenburg+92}, \citet{Brandenburg+93}, \citet{Moss+Sokoloff08}, \citet{Moss+10}, \citet{Beck+12}, \citet{Henriksen+Irwin16}, \citet{Shukurov+19}, \citet{Woodfinden+19}, and \citet{Henriksen+Irwin21}.

\subsubsection{The effect of starbursts on magnetic field}
In \textsc{galform}, 
when a starburst is triggered by a galaxy merger or disc instability,
gas is transferred from the disc to the spheroid \citepalias{Lacey+16}. 
In \textsc{magnetizer}, magnetic fields are only modeled for the disc component.
However, intense amplification of turbulent magnetic fields is likely to occur during a starburst.
For this reason, the magnetic field strengths and radio luminosities 
of some of the most actively star-forming galaxies in our model may be underestimated.

\subsubsection{The turbulent speed}\label{sec:vturb}
Our model does not include any variation of $v\turb$ within a given galaxy.
Future work should move toward a more physical (as opposed to empirical)
model for the turbulent speed that allows for its radial variation.
In general, the turbulence parameters, 
including $v\turb$ and the turbulent correlation length $l$, 
could be modelled to depend on more accessible properties
like the SFR surface density \citep[e.g.,][]{Chamandy+24,Nazareth+25}.

We note that the turbulent speed varies very significantly between the ISM phases, 
being higher in the hot gas than in the warm phase. 
Meanwhile, the fractional volumes of the warm and hot gas, as well as the turbulent speeds and magnetic fields in them, 
are sensitive to the SFR (Qazi et al., 2026, in preparation). 
Addressing these effects would require a better understanding of the generation mechanisms and parameters of magnetic fields 
and cosmic rays in the multi-phase ISM.

\subsubsection{Magnetic feedback}
While the role of the magnetic feedback on the parent galaxy remains poorly understood,
it may have significant effects on, for example, star formation
\citep[e.g.][]{Bogue+26}, the structure of the ISM and outflows 
\citep[e.g.][]{Evirgen+19,Martin-alvarez+26} and the multi-phase ISM structure (Qazi et al., 2026, in preparation). 
Unlike full MHD simulations, our model does not include the feedback of magnetic fields on processes like star formation 
that are handled by \textsc{galform}.
In \textsc{magnetizer}, magnetic feedback is included in the dynamical quenching non-linearity of the mean-field dynamo
as well as in the midplane pressure term affecting the structure of the gaseous disc.
To include self-consistently magnetic feedback in key galaxy formation processes, 
the \textsc{galform} and \textsc{magnetizer} components of our model would need to be merged into a single code; 
this is left for future work.

\subsubsection{Anisotropy of the small-scale magnetic field}\label{sec:anisotropy}
Some observational studies have argued that an anisotropic small-scale component of the magnetic field \citep{Jaffe+10}
is needed to explain synchrotron data in nearby galaxies,
and find that this component often dominates the polarized synchrotron emission \citep[][and references therein]{Beck+19}.
However, the situation is still far from clear, 
owing to various challenges in interpreting the data to construct a magnetic field model. 
Such anisotropy could be caused, for instance, by the turbulent tangling of the large-scale magnetic field $\ol{\bm{B}}$, 
which increases the small-scale magnetic field in the direction perpendicular to $\meanv{B}$,
and by the galactic differential rotation,
which stretches the turbulent cells in the azimuthal direction 
and increases the magnitude of $b_\phi$ relative to the other components
within the time-scale of the eddy turnover time $\tau$ \citep[e.g.,][]{Hollins+17}.
Including such effects might enhance the radio luminosity and strengthen the correlation between $\Lspec$ and $V\rot$.
In the interest of keeping the model as simple as possible, 
we have not included such effects, but we plan to explore this avenue in future work.

\section{Conclusions}\label{sec:conclusions}
We have developed a three-stage galaxy formation--galactic dynamo--synchrotron radiation
model to predict the radio emission from a large sample of galaxies (about $2\times10^5$)
as it evolves from a high redshift to $z=0$.
In this work, Paper~I of a series, we discuss the total synchrotron emission 
and its correlations with global galaxy parameters, including the star formation rate  
and the galactic rotation speed. 
Observational studies have found these quantities to be strongly correlated in nearby galaxies,
and we obtain similarly strong correlations in our model.
Our main conclusions can be summarized as follows:
\begin{enumerate}
    \item We compile observational data from multiple sources 
      and extend the correlation analysis of \citetalias{Tabatabaei+16}
      to include several more star-forming galaxies. 
      The empirical correlations identified by \citetalias{Tabatabaei+16} 
      between $\Lspec$ and both $\SFR$ and $V\rot$ remain highly significant in the larger data set.
      However, the correlation coefficients found for $\Lspec$--$V_{\rm rot}$ 
      for the combined data set are significantly lower than 
      for the smaller sample of \citetalias{Tabatabaei+16} (see Section~\ref{sec: T16_comp}).
    \item Our theoretical model reproduces the strong correlations between $\Lspec$ 
      and both $\SFR$ and $V\rot$ seen in observations of nearby galaxies,
      with best-fitting power laws that are in a good agreement with those inferred from the observational data, 
      and with enough dispersion to accommodate the scatter in the data.
    \item The degree of correlation in our model and associated power law exponents are frequency-independent,
      whereas the overall magnitude of the synchrotron luminosity is inversely related to the frequency, 
      in agreement with the observations. 
    \item These correlations are mainly driven by the small-scale magnetic field, 
      which dominates over the large-scale field at low redshifts.
      The rms strength of the small-scale field is parameterized as $b = f_b B\eq = f_b(4\uppi \rho v\turb^2)^{1/2}$,
      where $f_b=\const$ (we set $f_b=0.8$ in our fiducial model), 
      $B\eq$ is the field strength that corresponds to the energy equipartition with turbulence, 
      $\rho$ is the gas density, and $v\turb$ is the root-mean-square speed of the turbulent flow.
      Hence, the results are almost independent of the adjustable dynamo parameter $R_\kappa$, 
      which affects the large-scale magnetic field only.
    \item These correlations are also not very sensitive 
      to the 
      adjustable parameter, $f_b$.
      Values in the range $0.5\lesssim f_b \lesssim 1.0$ produce reasonable fits to the data.
      This suggests that the small-scale component of galactic magnetic fields
      is stronger than the saturation values predicted by fluctuation dynamo models 
      \citep[e.g.][]{Federrath+11}. 
      This may be caused by the large-scale shear \citep[e.g.,][]{Singh+17, Hollins+17} 
      and/or turbulent tangling of the large-scale magnetic field \citep[e.g.,][]{Gent+24}.
    \item Nor are the correlations sensitive to the particular ``flavor'' of the model 
      (i.e.,~\texttt{Fiducial} or \texttt{J24}).
      However, including \textit{all} the gas in $B\eq$ (as in \texttt{Fiducial})
      produces better agreement with observations than including the diffuse gas alone,
      as in \texttt{J24}.
    \item The degree of correlation is affected more strongly by $\rho$ than by $v\turb$. 
      However, as the synchrotron radiation is emitted by the entire galaxy, 
      the total synchrotron luminosity mainly depends on the total gas mass 
      rather than on the local or average gas density.
      Thus, we find that the $\Lspec$--$\SFR$ correlation is a consequence 
      of the well-known correlation between $\SFR$ and $M\gas$.
    \item The turbulent speed $v\turb$, which is assumed to be constant within each galaxy in our model
      but varies between galaxies in accordance with the SFR, 
      only affects the degree of correlation for a small fraction of galaxies with
      $\SFR > 1\Msunyr$ because in this regime $v\turb$ increases with $\SFR$, 
      whereas for smaller $\SFR$ we set $v\turb=\const$. 
      The increase in $v\turb$ at high $\SFR$ helps to explain the excess luminosity of high-SFR galaxies 
      as well as the  large variability in the luminosity between high-SFR galaxies.
    \item We also obtain a strong correlation between $\Lspec$ and $V\rot$, 
      consistent with observations, but we find that this correlation is strong only for 
      actively star-forming (main-sequence) spiral galaxies which have $\sSFR$ above a certain threshold 
      ($\sSFR\gtrsim 10^{-10.4}\rm yr^{-1}$ at $z=0$).
    \item The $\Lspec$--$V\rot$ correlation
      is a consequence of the correlation between $V\rot$ and $M_\star$ on the star-forming main sequence
      (the stellar mass Tully--Fisher relation).
      On the star-forming main sequence, $\SFR$ and $M_\star$ are tightly correlated, and since $\Lspec$ scales with $\SFR$, this naturally leads to a non-causal correlation between $\Lspec$ and $V\rot$.
    \item Our model predicts that these correlations persist up to at least redshift $z \simeq 3$,
      with roughly constant correlation coefficients and increasing slope (power law exponent).
      The increasing slope is caused by a strong dependence of the magnetic field strength
      on $\SFR$ at high $\SFR$ in our model.
      But for $z\gtrsim1$, 
      the situation becomes complicated and the results somewhat less reliable 
      because both the large-scale and small-scale magnetic field components are important
      (whereas the small-scale magnetic field dominates at low redshifts),
      and because certain parameters of the model 
      (the \textsc{magnetizer} parameters $f_b$ and $R_\kappa$ in addition to \textsc{galform} parameters) 
      have been calibrated using mainly low-redshift data.
\end{enumerate}
The model presented is rather stable to changes in parameter values, 
and the level of agreement with the observational data rather robust.
The present work lends some support to the common assumption that the magnetic field
is in approximate local energy equipartition with turbulence. 
Nevertheless, first-principles physics-based approaches are needed 
to ultimately eliminate the dependence of models on still crude assumptions and achieve deeper understanding.

\section*{Acknowledgements}
The authors are grateful to Kandaswamy Subramanian, Nishikanta Khandai, Tuhin Ghosh and Carlton Baugh for useful discussions, 
and to Fangxia An for providing the MIGHTEE-COSMOS data. We thank the
anonymous referee for their insightful and constructive comments that helped to improve this work. 
We acknowledge the use of the Pegasus high-performance computing cluster 
at the Inter-University Centre for Astronomy and Astrophysics (Pune, India). LC acknowledges support from the Department of Atomic Energy grant RIN-4001.
CJ gratefully acknowledges support from the RUSA 2.0 (T3A) project. 

\section*{Data and Code Availability}
Observational data are compiled from the literature and presented in Table~\ref{tab:data}.
Simulation outputs will be made available upon request to the authors.
Source codes for computing the luminosity and making the plots, along with a small subset of example inputs, are publicly available at \url{https://github.com/SUKANTAG285/MAGNETIZER-RC} and archived on Zenodo \cite{ghosh_2026_21840438}. 




\bibliography{reference}{}
\bibliographystyle{aasjournalv7}

\appendix

\section{The $Z$-component of the mean magnetic field}\label{app:Bz}
In our models, both $\mbr$ and $\mbp$ 
decrease exponentially in $Z$ with the scale height $2h\disc$, 
\[
  \mbr = \mbr'(r)\,\Exp{-|Z|/2h\disc}, \qquad
  \mbp = \mbp'(r)\,\Exp{-|Z|/2h\disc},
\]
where the prime denotes the midplane values.
From the divergence-free condition with azimuthal symmetry, we have
\begin{equation}
\begin{split}
  \frac{\partial B_Z}{\partial Z} &= -\frac{1}{r} \frac{\partial}{\partial r} (r B_r) \\
  &= - \left( \frac{B_r^{'}}{r} + B_r^{'}\frac{|Z|}{2h\disc^2(r)}\frac{d h\disc}{d r} +\frac{\partial B_r^{'}}{\partial r} \right)e^{-|Z|/2h\disc(r)} \\
  &\equiv \mathcal{I}.
\end{split}
\end{equation}
Therefore,
\begin{equation}
\begin{split}
  B_Z(r,Z) &= -\int_0^{Z} \mathcal{I}(r,Z') \,dZ'\\
   &= \frac{Z}{|Z|} \bigg\{ 2h\disc(r)\left(\frac{B'_r}{r} + \frac{\partial B'_r}{\partial r}\right)
  \left( \Exp{-|Z|/2h\disc(r)} -1 \right) \\
  &\qquad+ B'_r\,\frac{d h\disc}{d r}
  \left[\left( \frac{|Z|}{h\disc(r)} + 2\right) \Exp{-|Z|/2h\disc(r)} - 2 \right] \bigg\}.
\end{split}
\end{equation}

\section{The turbulent speed and star formation rate}\label{app:v_turb-SFR}
\begin{figure}
\centering
	\includegraphics[width=0.9\columnwidth]{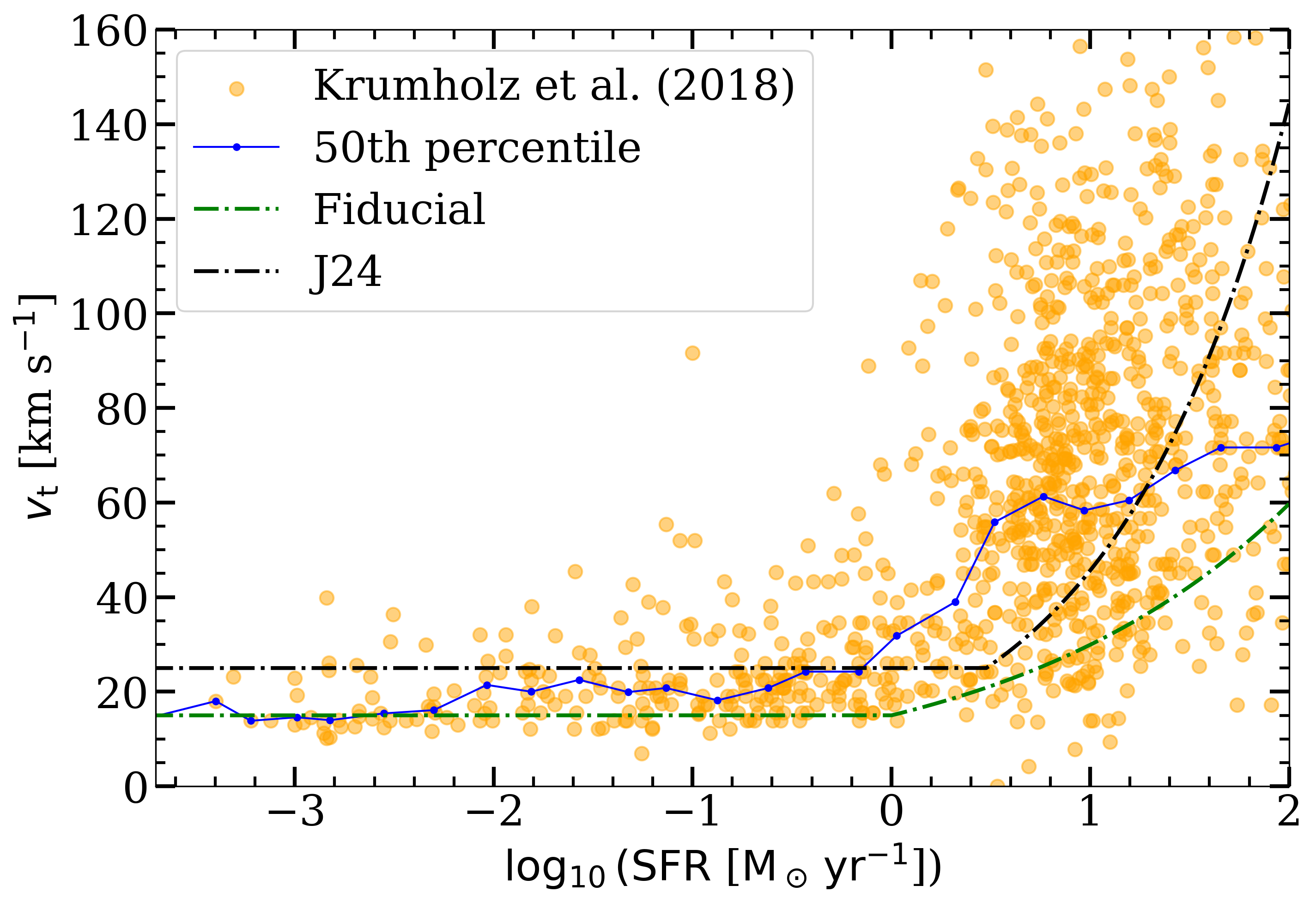}
    \caption{The scatter plot of the three-dimensional velocity dispersion as a function of the star formation rate 
    for the observational data from \citet[][and references therein]{Krumholz+18} (yellow circles). The blue curve with points represents the median of the scatter. 
    The green dash-dotted line represents the best-fitting median of the data of the form given in equation~\eqref{eq: sfr-vt} 
    (used in the \texttt{Fiducial}, \texttt{LS} and \texttt{SS} models), 
    whereas the black dash-dotted line represents the fitted relation of \citetalias{Jose+24}, 
    described by the same functional form as equation~\eqref{eq: sfr-vt}, 
    but with $\SFR_0 = 3\Msunyr$ and $c = 0.5$.}
    \label{fig:v_turb-SFR}
\end{figure}

Figure~\ref{fig:v_turb-SFR} 
shows the relation between the three-dimensional velocity dispersion 
and the global star formation rate of galaxies \citep{Krumholz+18} and the fits to these data used in our models. 

\section{Bulge-to-total mass ratio}\label{app:B_T_ratio}
\begin{figure}
\centering
	\includegraphics[width=0.85\columnwidth]{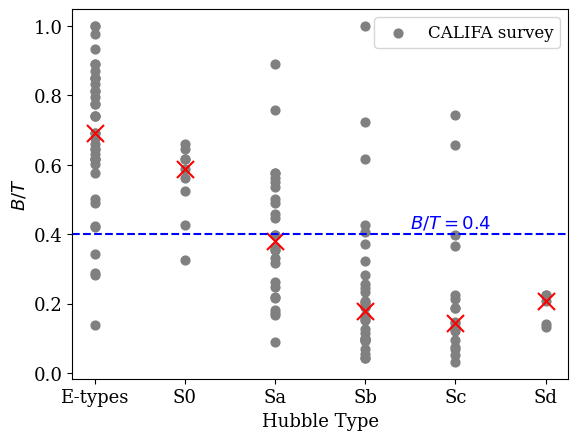}
    \caption{The bulge-to-total mass ratio $(B/T)$ 
    for different Hubble-type galaxies, using the CALIFA survey data from \citet{Mendez+21}. 
    Elliptical galaxies (E0–E7) are shown as E-types, 
    while lenticular/spiral galaxies are grouped as S0 (S0–S0a), Sa (Sa–Sab), Sb (Sb–Sbc), Sc (Sc–Scd), and Sd (Sd–Sdm). 
    Red crosses indicate the median $B/T$ values for each Hubble type.}
    \label{fig:Bulge_to_total_ratio}
\end{figure}

Figure~\ref{fig:Bulge_to_total_ratio} shows the stellar bulge-to-total mass ratio $(B/T)$ 
for different Hubble-type galaxies in the CALIFA survey. The data are taken from \citet{Mendez+21}, where the bulge and disc are separated by spectrophotometric decomposition and stellar masses are computed using the stellar population analysis for the bulge and disc separately. In the figure, galaxies are grouped as follows: ellipticals or E-types (E0–E7), S0 (S0–S0a), Sa (Sa–Sab), Sb (Sb–Sbc), Sc (Sc–Scd), and Sd (Sd–Sdm). The red crosses represent the median $B/T$ values for each Hubble type. The median $B/T$ decreases gradually from early- to late-type galaxies. For normal and late-type spiral galaxies (Sa–Sd), the median $B/T$ is typically $\lesssim 0.4$, whereas early types (E and S0) show median values $\gtrsim 0.6$. In this study, we use $B/T\le 0.4$ to select the star-forming disc galaxies. Note that the galaxies in this sample are unbarred. Barred galaxies could exhibit higher $B/T$ ratios.

\section{Rotation curves}\label{App: Flat rotation curve}
\begin{figure}
\centering
     \includegraphics [width=\columnwidth, clip=true, trim= 100 0 100 0]{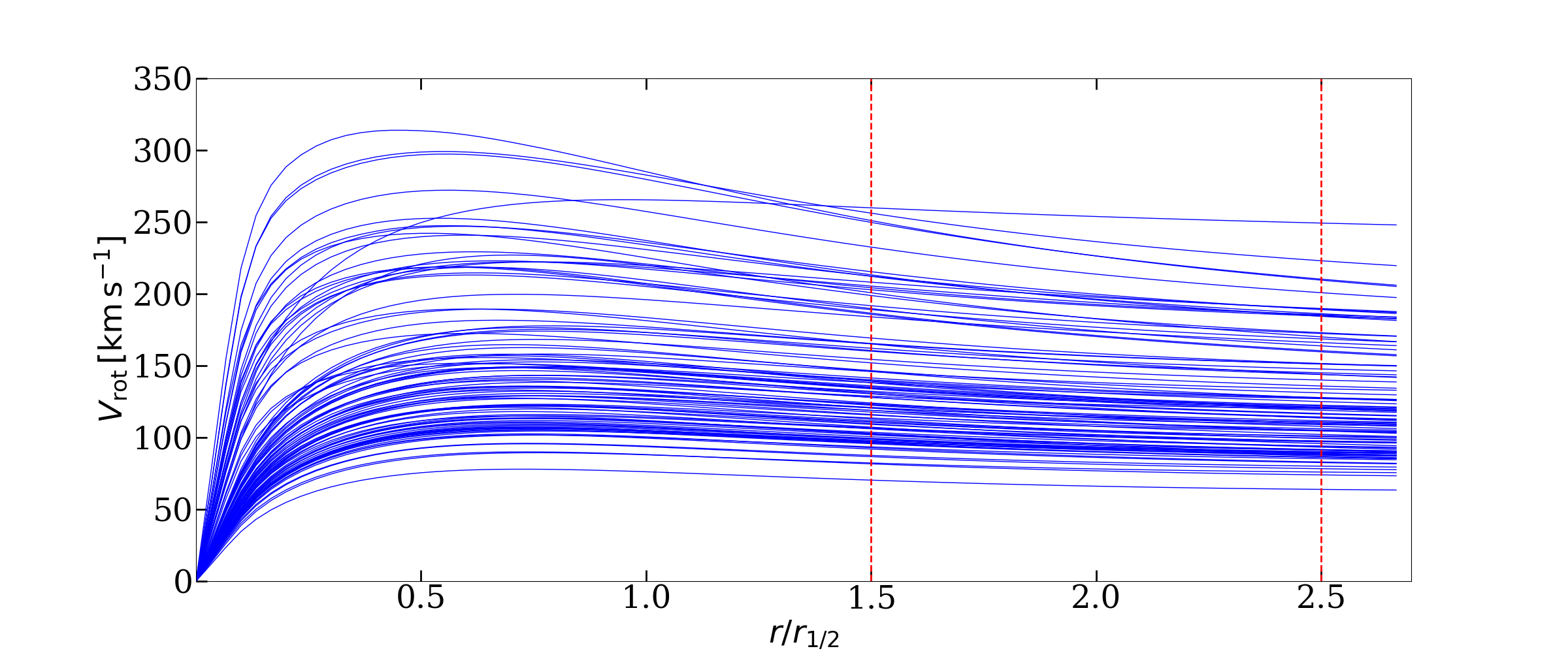}
     \caption{The rotation speed $V\rot$ as a function of the galactocentric radius 
     for $100$ randomly selected galaxies at redshifts $z= 0, 1$ and $2$. 
     The radius is normalized by the half-mass radius $r_{1/2}$.
     Vertical red lines are shown at $1.5\,r_{1/2}$ and $2.5\, r_{1/2}$,
     between which $V\rot$ is measured.}
     \label{fig:Vrot_r}
\end{figure}

In Fig.~\ref{fig:Vrot_r}, we plot a selection of the rotation curves of our simulated galaxies, with 
the region $1.5r_{1/2}$ - $2.5r_{1/2}$ highlighted to confirm that this part of the rotation curve is relatively flat.

\section{Specific star formation rate cutoff for active star\as{-}forming galaxies}\label{app:sSFR}
\begin{figure*}
\centering
    \includegraphics[width=0.8\textwidth]{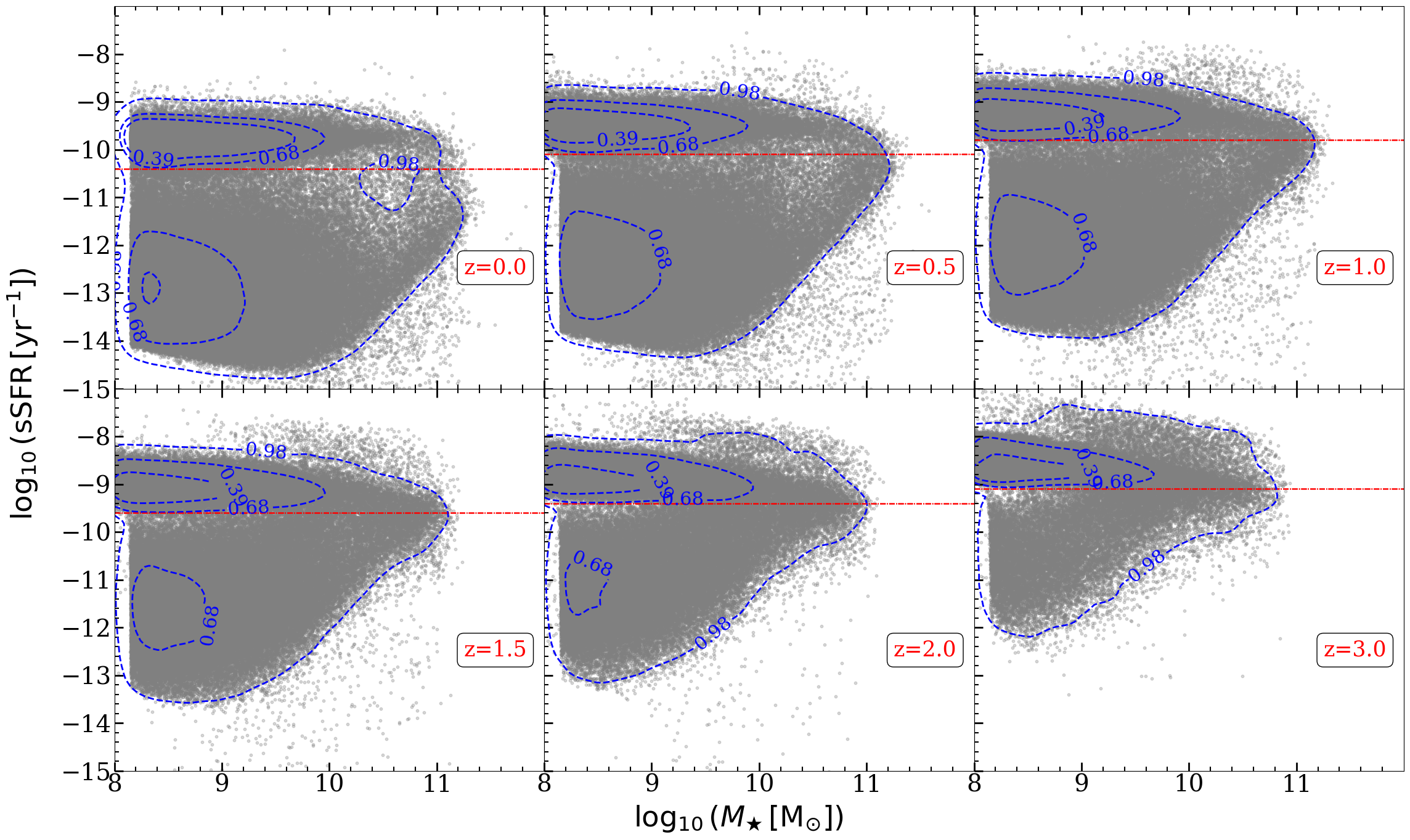}
    \caption{The redshift evolution of the specific star formation rate $\sSFR$ versus the stellar mass.
    The blue dashed contours represent the $39\%$, $68\%$ and $99\%$ confidence levels of the 2D kernel density estimate.
    Horizontal red dashed lines show the lower threshold sSFR 
    (set at the 68th percentile of the main-sequence population) values used to identify actively star-forming galaxies.}
    \label{fig:sSFR_Mstar}
\end{figure*}
The \textsc{galform} model produces a wide range of star-forming galaxies, 
spanning highly active systems to nearly passive ones. 
The actively star-forming galaxies exhibit a tight correlation between their SFR 
and stellar mass, commonly referred to as the star-forming main sequence. 
To separate the active and quiescent populations, 
we compute the sSFR, defined as the ratio of the SFR to the stellar mass.
Figure~\ref{fig:sSFR_Mstar} shows the redshift evolution ($z=0$--$3$) of $\sSFR$ as a function of the stellar mass. 
The horizontal red dashed lines mark the lower $\sSFR$ threshold used to identify actively star-forming galaxies. 
This threshold is defined as the 68th percentile of the main-sequence population at each redshift. 
At $z=0$, the lower threshold $\sSFR$ for active star-forming galaxies lies in the range $10^{-11}$--$10^{-10}\yr^{-1}$, 
consistent with observations as well as other galaxy formation models \citep{Katsianis+21a}. 
This threshold increases with the redshift \citep{Speagle+14}.

\section{Midplane gas pressure}\label{app: midplane pressure}

The pressure is the sum of the thermal pressure ($P_{\rm th}$), turbulent pressure ($P_{\rm turb}$), 
magnetic pressure ($P_{\rm mag}$) and cosmic ray pressure ($P_{\rm cr}$),
\begin{equation}
\begin{aligned}
  P &= P_{\rm th} + P_{\rm turb}  + P_{\rm mag} + P_{\rm cr} \\
    &= \rho_{\rm disc}\left[\frac{c_{\rm s}^2}{\gamma_{\rm ad}} 
       + \frac{1}{3}v_{\rm turb}^2(1+\xi+\xi\epsilon)\right] \\
    &\equiv \zeta \rho_{\rm disc} v_{\rm turb}^2,
\end{aligned}
\end{equation}
where $\gamma_{\rm ad}$, $\xi$ and $\epsilon$ are fixed parameters of the model.
Using
\begin{equation}
  b=f_b B\eq=f_b(4\uppi\rho\disc)^{1/2}v\turb, 
\end{equation}
the pressure due to the random component of the magnetic field is given by 
\begin{equation}
  P_b=\frac{b^2}{8\uppi}=\frac{f_b^2B\eq^2}{8\uppi}=\frac{f_b^2\rho\disc v\turb^2}{2}.
\end{equation}
If the magnetic pressure is dominated by this random component, we have
\begin{equation}
  P_{\rm mag} \approx P_b=\frac{f_b^2}{2}\rho\disc v\turb^2.
\end{equation}
Thus, we obtain
\begin{equation}\label{xi}
  \xi=\frac{3}{2}f_b^2.
\end{equation}
We choose $\epsilon=1$ to obtain
\begin{equation}\label{P}
  P \approx \rho_{\rm d}\left[\frac{c_{\rm s}^2}{\gamma_{\rm ad}}+\frac{1}{3}(1+3f_b^2)v_{\rm turb}^2\right],
\end{equation}
where $\gamma_{\rm ad}\approx5/3$. 
Note that our treatment neglects the pressure contribution from the large-scale magnetic field.

\section{Computing the thermal fraction}\label{app:thermal_frac}
The radio continuum spectrum can be written as
\begin{equation}\label{eq:E1}
  L_\nu^\mathrm{tot} = L_\nu^{\rm th} + L_\nu
  = A_1\,\nu^{-\alpha_{\rm th}} + A_2\,\nu^{-\alpha_{\rm nt}}. 
\end{equation}
Here, the thermal spectral index is $\alpha_{\rm th} \approx 0.1$ 
and the non-thermal spectral index $\alpha_{\rm nt}$ can have values in the range $0.7$ to $1.3$.
The thermal fraction at the frequency $\nu$ is written as,
\begin{equation}\label{eq:E2}
    f_\nu^{\rm th} =  \frac{L_\nu^{\rm th}}{L_{\nu}^\mathrm{tot}} 
    = \frac{L_\nu^{\rm th}}{L_\nu^{\rm th} + L_\nu}.
\end{equation}

Expressing $f_{\nu}^{\rm th}$ in terms of the thermal fraction $f_{\nu_0}^{\rm th}$ at a known frequency $\nu_0$ we have 
\begin{equation}\label{eq:E3}
    f_{\nu}^{\rm th} = \frac{L_\nu^{\rm th}/L_{\nu_0}^{\rm th}}{L_\nu^{\rm th}/L_{\nu_0}^{\rm th} 
    + L_\nu/L_{\nu_0}^{\rm th}} 
    = \frac{1}{1+ \left(1/f_{\nu_0}^{\rm th}-1\right)\left(\nu/\nu_0\right)^{\alpha_{\rm th}-\alpha_{\rm nt}}}.
\end{equation}
Using equation~\eqref{eq:E1} this leads to
\begin{equation}\label{eq:E5}
  \frac{L_\nu}{L_{\nu_0}^{\rm th}} 
  = \frac{L_\nu}{L_{\nu_0}}\frac{L_{\nu_0}}{L_{\nu_0}^{\rm th}} 
  = \frac{L_\nu}{L_{\nu_0}}\left(\frac{1}{f_{\nu_0}^{\rm th}}-1 \right).
\end{equation}

\section{Correlations for the \texttt{J24} model}\label{App: Alt_models}
\begin{figure*}[ht!]
    \centering
    \subfigure{
        \includegraphics[width=0.485\textwidth]{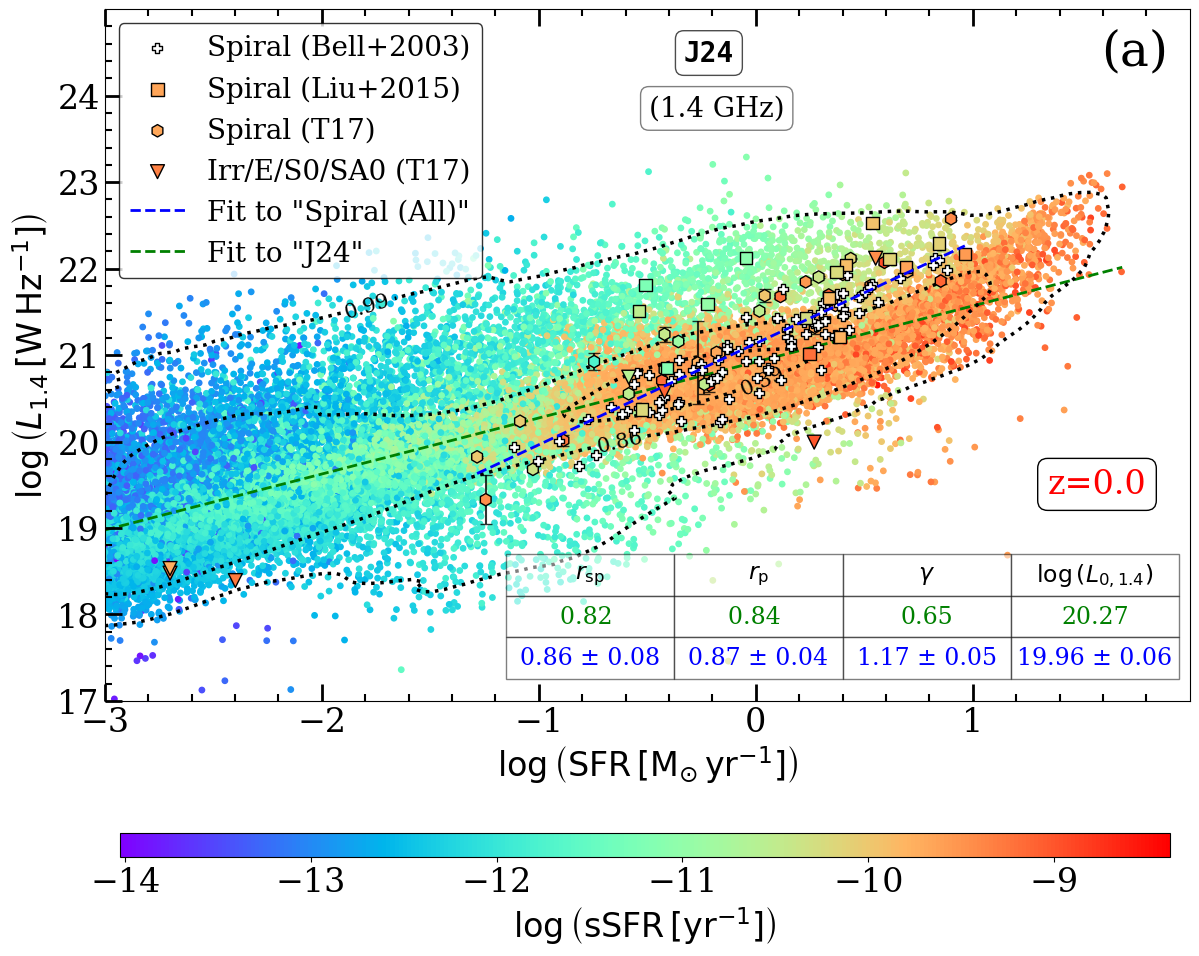}
        \label{fig: SI vs SFR 4.8 a}
    }
    \subfigure{
        \includegraphics[width=0.485\textwidth]{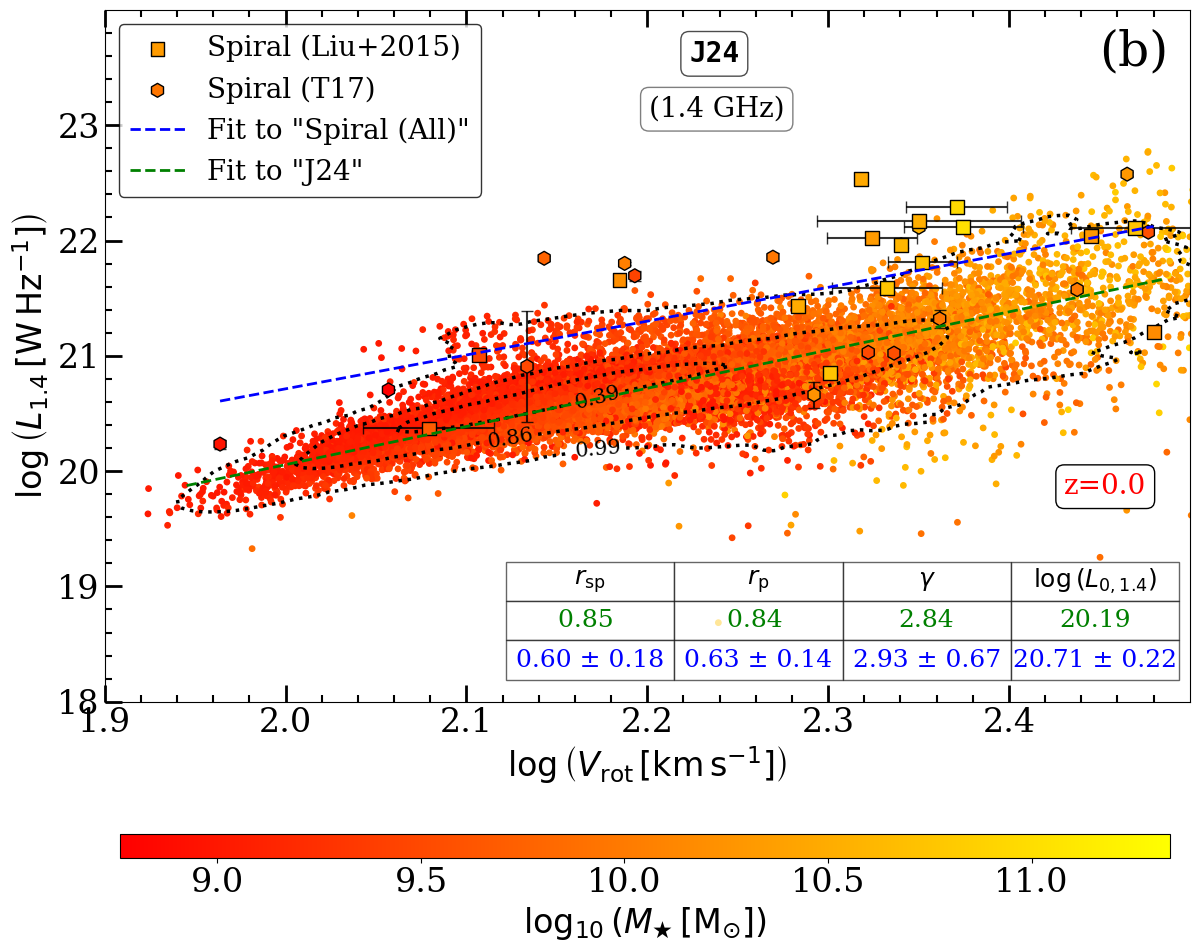}
        \label{fig:Lum_I_Vrot_4.8 a}
    }
    
    \caption{
    Similar to Figure~\ref{fig: SI vs SFR 4.8} but for \texttt{J24} model.
    }
    \label{fig: SI vs SFR 4.8 J24}
\end{figure*}
Figure~\ref{fig: SI vs SFR 4.8 J24} shows the correlations between the $1.4 \, \rm GHz$ rest-frame total synchrotron luminosity, $\Lspec$, and the star formation rate, SFR (Fig.~\ref{fig: SI vs SFR 4.8 a}), and between $\Lspec$ and the rotational velocity in the flat part of the rotation curve, $V\rot$ (Fig.~\ref{fig:Lum_I_Vrot_4.8 a}), for the \texttt{J24} model at $z=0$. The plotting conventions, symbols, fitting procedure, and inset tables are identical to those in Fig.~\ref{fig: SI vs SFR 4.8 b}.

\section{Contributions from $\meanv{B}$ and $\bfb$}\label{sec:relative}
\begin{figure}
    \centering
        \includegraphics[width=\columnwidth]{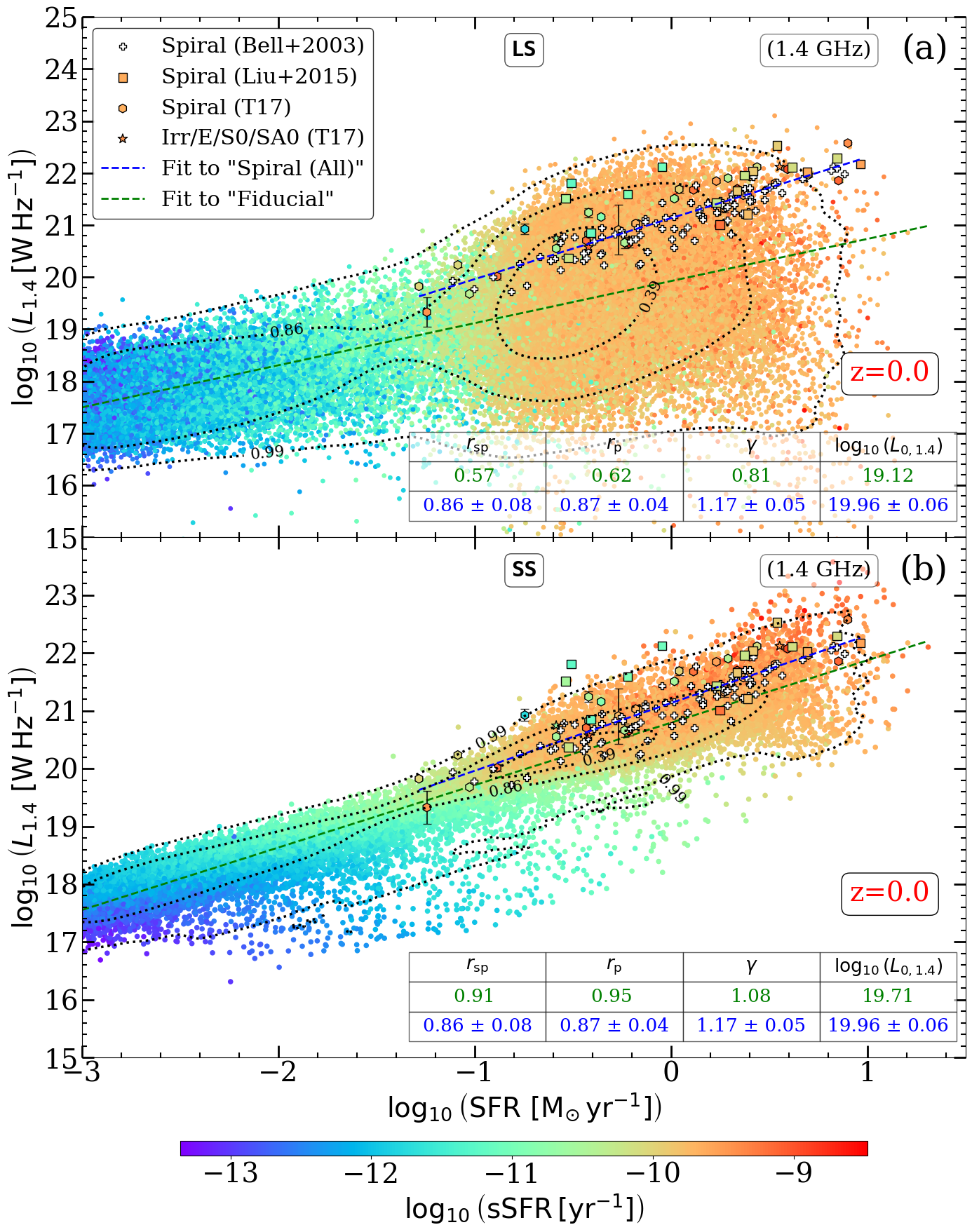}
        \label{fig:sub6a}
    \caption{Similar to Fig.~\ref{fig: SI vs SFR 4.8 b}, 
    but with the contributions of the large-scale ($\ol{B}$) (top panel) or small-scale ($b$) 
    (bottom panel) magnetic fields alone. 
    }
    \label{fig:L_SFR_LS_SS}
\end{figure}
\begin{figure}
    \centering
        \includegraphics[width=\columnwidth]{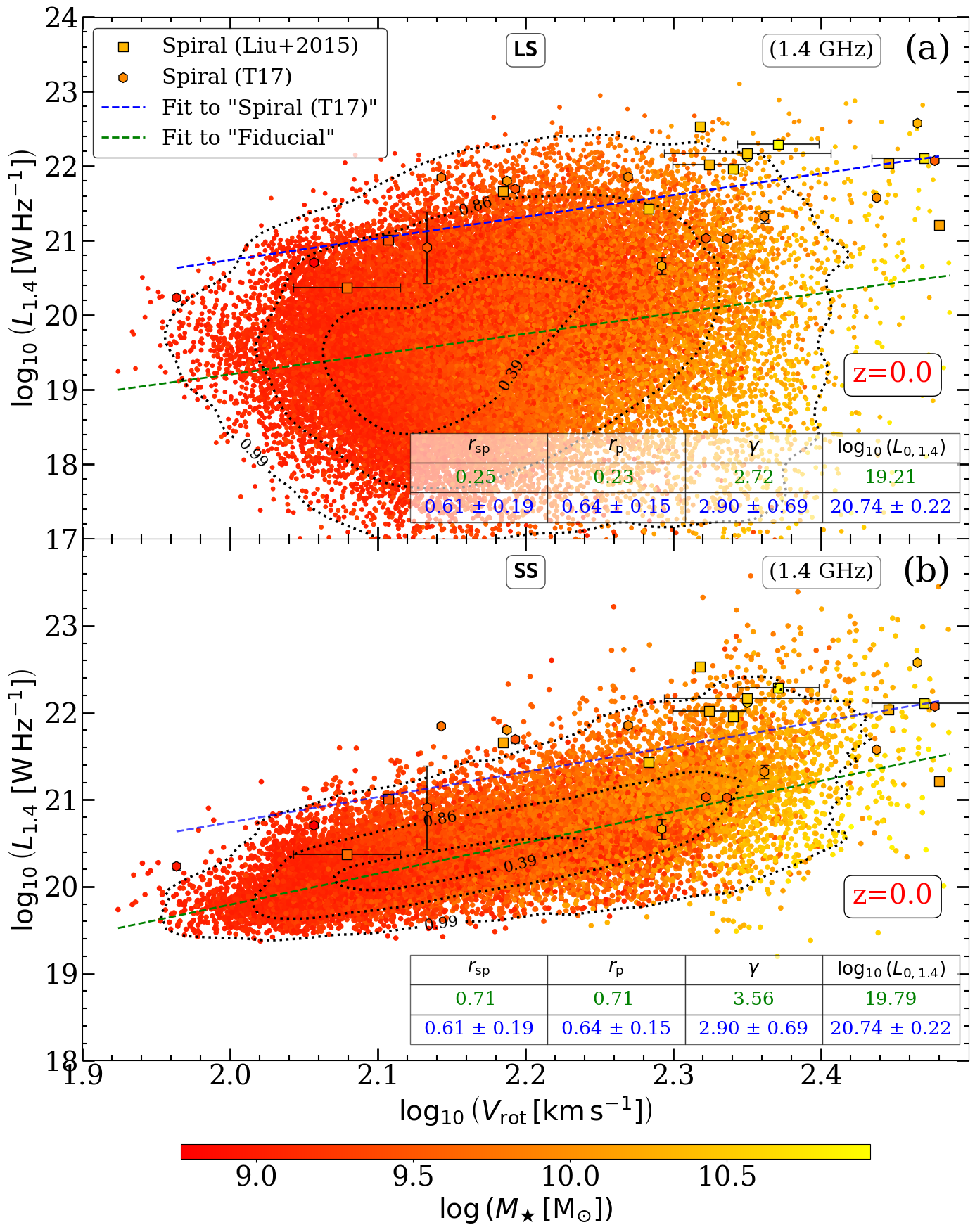}
        \label{fig:sub6b}
    \caption{Similar to Fig.~\ref{fig:Lum_I_Vrot_4.8 b}, 
    but with the contributions of the large-scale ($\ol{B}$) (top panel) or small-scale ($b$) 
    (bottom panel) magnetic fields alone. 
    }
    \label{fig:L_Vrot_LS_SS}
\end{figure}
Due to the local energy equipartition of magnetic fields and cosmic rays assumed in our model, 
the contributions of large-scale and small-scale magnetic fields are not strictly additive, so they cannot be separated (for details see Section~\ref{sec:relative_app}).
However, we can gain insight into the relative contributions of $\bfb$ and $\meanv{B}$ 
by calculating $\Lspec$ with one or the other set to zero.

Figure~\ref{fig:L_SFR_LS_SS} shows the correlation between $\Lspec$ and $\SFR$ at $\nu=1.4\GHz$ and $z =0$, with $b$ set to zero in Fig.~\ref{fig:L_SFR_LS_SS}a (Model~\texttt{LS})
and $\Bbar$ set to zero in Fig.~\ref{fig:L_SFR_LS_SS}b (Model~\texttt{SS}).
The correlation coefficients 
($r\Pe$ and $r\Sp$), 
slope, and overall luminosity magnitude agree much better with those of the observational data 
for Model~\texttt{SS} than for Model~\texttt{LS}.
Model~\texttt{SS} predicts a correlation that is slightly stronger, and the luminosity magnitude is a bit lower compared to the observational data and Model~\texttt{Fiducial}. This demonstrates that the large-scale magnetic field is not as important as the small-scale field 
for explaining the observed $\Lspec$--$\SFR$ correlation.
However, the large-scale field increases the scatter at high SFR 
and increases the overall luminosity somewhat, 
which helps to explain the brightest galaxies in $\Lspec$.

Figure \ref{fig:L_Vrot_LS_SS} presents the corresponding $\Lspec\text{--}V_{\rm rot}$ correlation. As in Fig.~\ref{fig:L_SFR_LS_SS}, the top panel (Model \texttt{LS}) includes only the large-scale magnetic field contribution, whereas the bottom panel (Model \texttt{SS}) includes only the small-scale magnetic field contribution. The results are qualitatively similar to those obtained for the $\Lspec-$SFR relation, with the small-scale magnetic field providing the dominant contribution to the correlation, while the large-scale magnetic field primarily enhances luminosity and scatter.


%



\end{document}